\documentclass[fleqn,twoside]{article}      
\usepackage{epsf,multicol,ifthen}
\usepackage{ujp}
\usepackage[cp1251]{inputenc}
\usepackage[english,ukrainian]{babel}
\usepackage{amstext}
\usepackage{amssymb,amsmath}
\usepackage{cite}

\usepackage{graphicx}

\usepackage{dcolumn}
\usepackage{bm}

\def\mys#1{\Sigma_{ } { }_{\bf #1}}
\def\bc{\begin{center}}
\def\ec{\end{center}}

\def\I2{{\large \rm \"{I}}}
\def\i2{\mbox{\footnotesize\rm \"l\hspace*{-2.45pt}l}}

\def\bi2{\mbox{\footnotesize\rm \bf \"l\hspace*{-2.75pt}l}}

\begin{document}

\newfont{\cyrfnt}{wncyr10 scaled 1200}

\newfont{\cyrB}{wncyb10 scaled 1440}

\newfont{\cyrI}{wncyi10 scaled 1440}

\mathindent=0pt%

\nazva{EXACTLY SOLVABLE MODELS: THE ROAD TOWARDS A RIGOROUS
TREATMENT OF PHASE TRANSITIONS IN FINITE NUCLEAR SYSTEMS
}  

\udk{539.12}

\nazvacol{Exactly Solvable  Models}%

\avtor{K.A. Bugaev$^{1,2}$, P. T. Reuter$^{3}$}%
\avtorcol{K.A. Bugaev, P. T. Reuter}%

\inst{\hspace*{-0.14cm}$^1$Bogolyubov Institute for Theoretical Physics, Nat. Acad. Sci. of Ukraine}%
\adr{(14$^b$, Metrologichna  Str., Kyiv 03143, Ukraine; e-mail: KABugaev@th.physik.uni-frankfurt.de)}%
\insti{Lawrence Berkeley National Laboratory}%
\adri{(1, Cyclotron Rd., Berkeley, CA 94720, USA )}%

\instii{Triumf, Canada}%
\adrii{(4004 Wesbrook Mall, Vancouver, Canada V6T 2A3;  e-mail: reuter@triumf.ca)}%

                
\setcounter{page}{489}%
\maketitl                 

\date{\today}

\anot{
We discuss exact analytical solutions of a variety of statistical models recently  obtained for finite systems  by 
a novel powerful mathematical method, the Laplace-Fourier transform.  
Among them are 
a constrained version of the  statistical multifragmentation model,
the Gas of Bags Model
and the Hills and Dales Model of surface partition. 
Thus, the Laplace-Fourier transform allows one  to  study the nuclear matter equation of state,
the equation of state of hadronic and quark gluon matter and surface partitions   on the same footing.
A complete analysis of the isobaric partition singularities of these models  is done for finite systems. 
The developed  formalism allows us, for the first time, to exactly define the finite volume analogs
of gaseous, liquid and mixed  phases of these models  from the first principles of statistical mechanics and demonstrate the pitfalls of  earlier  works. 
The found solutions may be used for  building  up  a  new  theoretical apparatus  
to rigorously study  phase transitions in finite systems. 
The strategic directions of  future  research   opened by these exact results are also  discussed.  
}



\vspace*{-0.7cm}

\hfill\begin{minipage}[t]{6.3cm}
{\it There is always a sufficient amount  of facts. Imagination is  what we lack.}\\
\hspace*{2.4cm} D. I. Blokhintsev 
\end{minipage}

\section{Theoretical  Description of   Phase Transitions in Finite Systems}

\vspace*{-0.2cm}

A  rigorous theory of critical phenomena in finite systems was not built up to now. 
However, the experimental studies of phase transitions (PTs) in some systems demand
the formulation  of such a theory. In particular, the investigations  of the  nuclear liquid-gas PT 
\cite{Bondorf:95, Gross:97, Moretto:97}
require the development of theoretical approaches which  would allow us to study
the  critical phenomena without going into the thermodynamic limit  $V \rightarrow \infty$ ($V$ is 
the volume of the system) because  such a limit does not exist due the long range Coulomb interaction.
Therefore, there is  a great need in the theoretical approaches which may shed light on the ``internal mechanism''  of how the PTs happen in finite systems.

The general situation in the theory of  critical phenomena for finite (small) 
systems  is not very optimistic  at the moment because  
theoretical  progress in this field has been slow.
It is well known that
the mathematical theory of phase transitions was worked out by 
T. D. Lee and C. N. Yang \cite{LeeYang}. 
Unfortunately, there is no direct generic  relation between the physical observables and zeros of the grand canonical partition  in a complex fugacity  plane.
Therefore, we know  very well what are the  gaseous phase and liquid at infinite volumes: 
mixture of fragments of all sizes and ocean, respectively.  
This  is known both for pure phases and for their mixture, but, 
despite some limited  success \cite{Chomaz:03},  
this general approach is not useful for  the specific problems of critical phenomena in  finite systems
(see Sect. VIII below). 

The tremendous complexity  of   critical  phenomena in finite systems prevented 
 their  systematic and rigorous  theoretical  study.
For instance, even the best   formulation of  the statistical mechanics and 
thermodynamics of finite systems  by Hill \cite{Hill} is not rigorous while discussing PTs.
As a result,
the absence of  a  well established definition of the liquid and mixed phase for finite volumes
delays the progress of several related fields, including  the theoretical and  experimental 
searches for 
the reliable 
signals  of  several  PTs which are expected to exist in  strongly interacting matter. 
Therefore, {\it  the  task of  highest priority} of the  theory of  critical phenomena
 is  {\it to define  the  finite volume analogs of phases  
from first principles of statistical mechanics.}   
{ At present  it is unclear  whether such definitions 
can be made for  a  general case, but it turns out that such finite volume definitions 
can be formulated for a variety of  realistic nonclassical (= non mean-field)  statistical models  which are successfully 
used in nuclear multifragmentation and in relativistic heavy collsisions. }

About 25 years ago, when the theoretical  foundations
of nuclear multifragmentation  were established,
there was an illusion that the theoretical basis is simple and clear and, therefore,
we need only the data and  models which will describe them.  
The  analysis of finite volume systems has proven to be very difficult. 
However, there was a clear way out of troubles by  making numerical codes 
that are able to describe the data.  
This is, of course, a common way 
to handle such problems and there were many successes achieved in this way 
\cite{Bondorf:95, Gross:97, Moretto:97,Randrup:04}.
However, there is another side of the coin which tells us that our understanding did 
not change much since then. This is so
because the numerical simulations  of this level do not provide us with any proof.  
At best they just demonstrate something.
With  time the number of codes increased, but the common theoretical approach was not developed.  This led to a bitter result - there are many good guesses in the nuclear multifragmentation community, 
but, unfortunately,  little analytical work to back up these expectations.
As a result  the absence of a firm theoretical ground led to formulation of such  highly speculative 
``signals'' of  the  nuclear liquid-vapor PT  as negative heat capacity \cite{Negheat:1, Negheat:2}, bimodality
\cite{Bimodality:1}, 
which later on were disproved, in Refs \cite{Negheat:3} and \cite{Bimodality:2}, respectively. 

{
Thus,  there is  a paradoxic   situation: there are many experimental data and 
facts, but there is no a single theoretical  approach which is able to describe them. 
Similar to the  searches for quark-gluon plasma (QGP) \cite{QM:04} there is  
lack of a firm and rigorous theoretical approach
to  describe phase transitions in finite systems. 
}

However, 
our understanding of the multifragmentation phenomenon 
\cite{Bondorf:95, Gross:97, Moretto:97}  was  improved  recently,  when 
an exact analytical solution of a simplified version of 
the statistical multifragmentation model (SMM) \cite{Gupta:98,Gupta:99} 
was found in Refs. \cite{Bugaev:00,Bugaev:01}.  
These  analytical results not only allowed us to understand the important
role of the Fisher exponent  $\tau$  on the phase structure  of  the nuclear 
liquid-gas PT and the properties of  its (tri)critical point,
but  to  calculate the critical indices  $\alpha^\prime, \beta,\gamma^\prime, \delta$ of the SMM \cite{Reuter:01}  as  functions of index  $\tau$.
 The determination of the simplified  SMM  exponents  allowed us to 
show  explicitly \cite{Reuter:01} that, in contrast to  expectations,
{\it the scaling relations for critical indices of the SMM
differ from the corresponding relations of  a well known Fisher  droplet model (FDM)  \cite{Fisher:67}.}
This exact analytical solution allowed us to  predict
a narrow range of values, $1.799< \tau < 1.846$,
which, in contrast to FDM value $\tau_{FDM}  \approx 2.16$,  
is consistent with ISiS Collaboration data  \cite{ISIS}
and
EOS Collaboration data  \cite{EOS:00}. 
This finding  is  not only of a  principal theoretical  importance, since it allows one to find out the universality
class of the nuclear liquid-gas phase transition, if  $\tau$  index  can be determined from 
experimental mass distribution of fragments, but also  it   
enhanced  a great activity  in extracting the value of  $\tau$  exponent from 
the data \cite{Karnaukhov:tau}.

It is necessary to stress  that  such  results {\it in principle} cannot be obtained  
either  within  the widely used 
mean-filed approach or numerically.  
This is the reason why exactly solvable models with
phase transitions play a special role in  statistical
mechanics - they are the benchmarks of our understanding of critical phenomena that occur  in 
more complicated  substances. 
They are our theoretical laboratories, where we can study the most fundamental problems of critical phenomena which cannot be studied  elsewhere. 
Their  great  advantage compared to other methods  is that they provide us with 
the information obtained directly  from the first principles of statistical mechanics being 
unspoiled by mean-field or other simplifying approximations without which the analytical 
analysis is usually impossible.  On the other hand an exact analytical solution gives the physical
picture  of  PT, which cannot be obtained by numerical evaluation.  Therefore, one can expect 
that an extension of the exact analytical solutions to finite systems may provide us with  the 
ultimate and reliable experimental signals of the nuclear liquid-vapor PT  which are established
on a firm theoretical ground of statistical mechanics. This, however, is a very difficult  general task
of the critical phenomena theory in finite systems.

Fortunately,  we do not need to solve this very general task, but to find  its solution for  a specific 
problem of nuclear liquid-gas PT, which is less complicated and more realistic. In this case
the straightforward way is to start from a few statistical models, like FDM and/or SMM, which are successful  in describing the most of the experimental data.  A systematic study of the various
modifications of the FDM for finite volumes was performed by  
Moretto and collaborators \cite{Elliott:05wci}  and 
it led to a discovery of thermal reducibility  of the fragment  charge spectra  \cite{Moretto:97}, 
to a determination  of  a quantitative liquid-vapor phase diagram containing the coexistence line
 up to critical temperature for small systems \cite{Elliott:02,Elliott:03},  to the generalization  of the FDM 
 for  finite systems and to a  formulation of the complement concept 
  \cite{Precomplement,Complement} 
 which allows one  to account for  finite size effects of (small)  liquid drop on the properties 
 of its vapor. 
  However,  such a systematic  analysis for the  SMM was  not  possible  until recently, when 
  its finite volume  analytical solution was found in \cite{Bugaev:04a}.

An invention of a new powerful mathematical method \cite{Bugaev:04a}, 
the Laplace-Fourier transform,  is a major theoretical breakthrough in the statistical mechanics
of finite systems of the  last decade  because it  
allowed us   to solve exactly  not only  the simplified SMM  for finite volumes 
\cite{Bugaev:04a}, but  also   a variety of statistical surface partitions  for  finite clusters  \cite{Bugaev:04b} 
 and to find out   their surface entropy   and to  shed light
 on a source of the Fisher exponent $\tau$.
It was shown \cite{Bugaev:04a}
that for finite volumes the analysis of the grand canonical partition (GCP) of the simplified SMM
is reduced to the analysis of the simple poles of the corresponding isobaric partition,  obtained 
as a Laplace-Fourier transform of the GCP.   Such a  representation of the GCP allows one
not only to show from first principles  that for finite systems there exist the complex 
values of the effective chemical potential, but 
to
define  the finite volume analogs of phases straightforwardly.  Moreover,  this method 
allows one to include into consideration all complicated features of the interaction (including the Coulomb one) which have been
neglected in the simplified SMM because it  was originally  formulated for infinite nuclear matter. 
Consequently,   the Laplace-Fourier transform  method opens  a principally new  possibility 
to study the nuclear liquid-gas phase transition directly from the partition of finite system 
without  taking its thermodynamic limit.  Now this method is also applied  \cite{Bugaev:05} to the finite
volume formulation  of the  Gas of Bags Model (GBM)   \cite{Goren:81} which is used to  describe the 
PT between the hadronic matter and  QGP. 
Thus,  the Laplace-Fourier transform method not only  gives an analytical  solution for a variety of 
statistical models  with PTs in finite volumes, but 
provides us with  a common framework   for several critical phenomena  in  strongly interacting matter. Therefore, it turns out that further applications and developments of this method are 
very promising and important  not only for the  nuclear multifragmentation community, but 
for several communities studying PTs in finite systems  because this method  may provide them with the firm theoretical foundations and a  common theoretical language.

It is necessary to remember that  further  progress  of this approach  and  its extension to other communities  cannot  be successfully  achieved  without new theoretical ideas about formalism it-self and its applications to the data measured in  low and high energy nuclear collisions. Both of these  require   essential  and coherent 
efforts of  two or three theoretical groups working on the theory of  PTs in finite systems, 
which, according to
our best  knowledge,  do not exist  at the moment either in  multifragmentation community or elsewhere. Therefore, {\it the second task  of  highest priority  is to attract young and promising
theoretical students} to these theoretical problems and create the necessary manpower to solve 
the up coming problems. Otherwise the negative 
consequences 
of a complete  dominance of experimental groups and  numerical codes will never be 
overcome  and  a  good chance to build up  a common  theoretical apparatus for  a few
PTs  
will be lost forever.  If this will be the case, then  an essential part  of the nuclear physics 
associated with nuclear multifragmentation will   have no chance  to survive in the next years.

Therefore,  the first  necessary step to resolve these two tasks of highest priority is to formulate 
the up to day achievements of the exactly solvable models and to discuss the strategy for their  further
developments and improvements  along with their possible impact on transport and hydrodynamic 
approaches. For these reasons 
the paper is organized as follows:  in Sect.  II we  formulate the simplified SMM and  present 
its analytical solution in thermodynamic limit; in Sect. III we  discuss the necessary conditions for 
PT of given order  and their relation to the singularities of the isobaric partition and apply these 
findings to the simplified SMM;  Sect. IV is devoted to the SMM critical indices  as  the functions 
of Fisher exponent $\tau$ and  their scaling relations;  the Laplace-Fourier  transform method  
is presented in Sect. V along with 
an exact analytical solution of the 
simplified SMM  which has a constraint  on  the  size of largest fragment,
whereas the analysis of  its isobaric partition singularities and the meaning of the complex values
of free energy  are  given in Sect. VI; 
Sect. VII and VIII are devoted to the discussion of the case without PT  and with it, respectively;
at the end of Sect. VIII there is a discussion 
of the Chomaz and Gulminelli's approach to bimodality  \cite{Chomaz:03};
in Sect. IX  we discuss 
the finite volume modifications of the Gas of Bags, i.e. the statistical model describing 
the PT between hadrons and QGP, whereas in  Sect. X  we formulate the Hills and Dales Model
for the surface partition and present  the limit of the vanishing amplitudes of deformations;
and, finally,  in  Sect. XI  we discuss the strategy of  future research which is necessary 
to build up a truly microscopic kinetics of phase transitions in finite systems.


\vspace*{-0.35cm}

\section{Statistical Multifragmentation  in Thermodynamic Limit}

\vspace*{-0.2cm}

The system states in the SMM are specified by the multiplicity
sets  $\{n_k\}$
($n_k=0,1,2,...$) of $k$-nucleon fragments.
The partition function of a single fragment with $k$ nucleons is
\cite{Bondorf:95}:
$
V \phi_k (T) = V\left(m T k/2\pi\right)^{3/2}~z_k~
$,
where $k=1,2,...,A$ ($A$ is the total number of nucleons
in the system), $V$ and $T$ are, respectively, the  volume
and the temperature of the system,
$m$ is the nucleon mass.
The first two factors  on the right hand side (r.h.s.) 
of 
the single fragment partition 
originate from the non-relativistic thermal motion
and the last factor,
 $z_k$, represents the intrinsic partition function of the
$k$-nucleon fragment. Therefore, the function $\phi_k (T)$ is a phase space
density of the k-nucleon fragment. 
For \mbox{$k=1$} (nucleon) we take $z_1=4$
(4 internal spin-isospin states)
and for fragments with $k>1$ we use the expression motivated by the
liquid drop model (see details in \mbox{Ref. \cite{Bondorf:95}):}
$
z_k=\exp(-f_k/T),
$ with fragment free energy
\begin{equation}\label{one}
f_k = - W(T)~k 
+ \sigma (T)~ k^{2/3}+ (\tau + 3/2) T\ln k~,
\end{equation}
with $W(T) = W_{\rm o} + T^2/\epsilon_{\rm o}$.
Here $W_{\rm o}=16$~MeV is the bulk binding energy per nucleon.
$T^2/\epsilon_{\rm o}$ is the contribution of
the excited states taken in the Fermi-gas
approximation ($\epsilon_{\rm o}=16$~MeV). $\sigma (T)$ is the
temperature dependent surface tension parameterized
in the following relation:
$
\sigma (T) = \sigma (T)|_{SMM}  \equiv \sigma_{\rm o}
[(T_c^2~-~T^2)/(T_c^2~+~T^2)]^{5/4},
$
with $\sigma_{\rm o}=18$~MeV and $T_c=18$~MeV ($\sigma=0$
at $T \ge T_c$). The last contribution in Eq.~(\ref{one}) involves the famous Fisher's term with
dimensionless parameter
$\tau$.  
As we will show later, at the critical (tricritical) point the fragment mass distribution 
will lose it exponential form and will become a power law $k^{-\tau}$.

{ It is necessary to stress that  the  SMM parametrization  of  the surface tension coefficient is 
not   a unique one. For instance, the FDM successfully  employs  another one
$\sigma (T)|_{\rm FDM}  = \sigma_{\rm o} [ 1~ - ~T/T_c].$ 
As we shall see in  Sect. IV  the temperature dependence of the surface tension 
coefficient  in the vicinity of the critical point will define the critical indices of the model, 
but   the following mathematical  analysis of the SMM  is general and  is valid for an arbitrary
$\sigma (T)$ function. 
}

The canonical partition function (CPF) of nuclear
fragments in the SMM
has the following form:
\begin{equation} \label{two}
\hspace*{-0.2cm}Z^{id}_A(V,T)=\sum_{\{n_k\}} \biggl[
\prod_{k=1}^{A}\frac{\left[V~\phi_k(T) \right]^{n_k}}{n_k!} \biggr] 
{\textstyle \delta(A-\sum_k kn_k)}\,.
\end{equation}
In Eq. (\ref{two}) the nuclear fragments are treated as point-like objects.
However, these fragments have non-zero proper volumes and
they should not overlap
in the coordinate space.
In the excluded volume (Van der
Waals) approximation
this is achieved
by substituting
the total volume $V$
in Eq. (\ref{two}) by the free (available) volume
$V_f\equiv V-b\sum_k kn_k$, where
$b=1/\rho_{{\rm o}}$
($\rho_{{\rm o}}=0.16$~fm$^{-3}$ is the normal nuclear density).
Therefore, the corrected CPF becomes:
$
Z_A(V,T)=Z^{id}_A(V-bA,T)
$.
The SMM defined by Eq. (\ref{two})
was studied numerically in Refs. \cite{Gupta:98,Gupta:99}.
This is a simplified version of the SMM, since  the symmetry and
Coulomb contributions are neglected.
However, its investigation
appears to be of  principal importance
for studies of the nuclear  liquid-gas phase transition.

The calculation of $Z_A(V,T)$  
is difficult due to  the constraint $\sum_k kn_k =A$.
This difficulty can be partly avoided by 
evaluating
the grand canonical partition (GCP) 
\begin{equation}\label{three}
{\cal Z}(V,T,\mu)~\equiv~\sum_{A=0}^{\infty}
\exp\left({\textstyle \frac{\mu A}{T} }\right)
Z_A(V,T)~\Theta (V-bA) ~,
\end{equation}
where $\mu$ denotes a chemical potential.
The calculation of ${\cal Z}$  is still rather
difficult. The summation over $\{n_k\}$ sets
in $Z_A$ cannot be performed analytically because of
additional $A$-dependence
in the free volume $V_f$ and the restriction
$V_f>0 $.
The presence of the theta-function in the GCP (\ref{three}) guarantees
that only configurations with positive value of the free volume 
are counted. However,
similarly to the delta function restriction in Eq.~(\ref{two}),
it makes again
the calculation of ${\cal Z}(V,T,\mu)$ (\ref{three}) to be rather
difficult. This problem  was  resolved   \cite{Bugaev:00,Bugaev:01} 
by performing the Laplace
transformation of ${\cal Z}(V,T,\mu)$. This introduces the so-called
isobaric
partition function (IP)  \cite{Goren:81}:
\begin{eqnarray} \label{four}
\hat{\cal Z}(s,T,\mu)~\equiv ~\int_0^{\infty}dV~{\textstyle e^{-sV}}
~{\cal Z}(V,T,\mu)
&=&
\hspace*{-0.0cm}\int_0^{\infty}\hspace*{-0.2cm}dV^{\prime}~{\textstyle e^{-sV^{\prime} } }
\sum_{\{n_k\}}\hspace*{-0.1cm}\prod_{k}~\frac{1}{n_k!}~\left\{V^{\prime}~\phi_k(T)~
{\textstyle e^{\frac{(\mu  - sbT)k}{T} }}\right\}^{n_k} \nonumber \\
&=&\hspace*{-0.0cm}\int_0^{\infty}\hspace*{-0.2cm}dV^{\prime}
~{\textstyle e^{-sV^{\prime} } }
\exp\left\{V^{\prime}\sum_{k=1}^{\infty}\phi_k ~
{\textstyle e^{\frac{(\mu  - sbT)k}{T} }}\right\}~.
\end{eqnarray}

\vspace*{-0.1cm}

\noindent
After changing the integration variable $V \rightarrow V^{\prime}$,
the constraint of $\Theta$-function has disappeared.
Then all $n_k$ were summed independently leading to the exponential function.
Now the integration over $V^{\prime}$ in Eq.~(\ref{four})
can be  done resulting in
\begin{equation}\label{five} 
\hat{\cal Z}(s,T,\mu)~=~\frac{1}{s~-~{\cal F}(s,T,\mu)}~,
\end{equation}

\vspace*{-0.3cm}

\noindent
where

\vspace*{-0.5cm}

\begin{eqnarray} \label{Osix}
& &\hspace*{-0.9cm}
{\cal F}(s,T,\mu) = \sum_{k=1}^{\infty}\phi_k ~ 
\exp\left[\frac{(\mu - sbT)k}{T}\right] =
 \left( \frac{mT }{2\pi}\right)^{\frac{3}{2} } 
\left[z_1 \exp\left(\frac{\mu-sbT}{T}\right) + 
\sum_{k=2}^{\infty}
k^{ -\tau} \exp\left(
\frac{(\tilde\mu - sbT)k -
\sigma k^{2/3}}{T}\right)\right]. 
\end{eqnarray}
%
Here we have introduced the shifted chemical potential
$\tilde{\mu}~\equiv~\mu ~+~W (T) $.
From the definition of pressure in the grand canonical ensemble
it follows that, in the thermodynamic limit,
the GCP of the system  behaves as 
\begin{equation}\label{gcpfunc}
p(T,\mu)~\equiv~ T~\lim_{V\rightarrow \infty}\frac{\ln~{\cal Z}(V,T,\mu)}
{V}
\quad \Rightarrow \quad 
{\cal Z}(V,T,\mu)\biggl|_{V \rightarrow \infty } \sim 
\exp\left[\frac{p(T,\mu)V}{T} \right]~. \biggr.
\end{equation}  
An exponentially over $V$ increasing part of ${\cal Z}(V,T,\mu)$
in the right-hand side of Eq.~(\ref{gcpfunc}) generates
the rightmost singularity $s^*$ of the function
$\hat{\cal Z}(s,T,\mu)$, because for $s<p(T,\mu)/T$ the
$V$-integral for $\hat{\cal Z}(s,T,\mu)$ (\ref{four}) diverges at its upper
limit. Therefore, 
in the thermodynamic limit, $V\rightarrow \infty$ the system pressure
is defined by this rightmost  singularity, $s^*(T,\mu)$, of  
IP $\hat{\cal Z}(s,T,\mu)$ (\ref{four}): 
\begin{equation}\label{ptmuii}
p(T,\mu)~=~T~s^*(T,\mu)~.
\end{equation}
Note that this simple connection of the rightmost \mbox{$s$-singularity}
of $\hat{\cal Z}$, Eq.~(\ref{four}), to the asymptotic,
 $V\rightarrow\infty$,
behavior of ${\cal Z}$, Eq.~(\ref{gcpfunc}), is a general mathematical
property
of the Laplace transform. Due to this property the study of the 
system behavior in the thermodynamic limit
$V\rightarrow \infty$ can be reduced to the investigation of
the singularities of $\hat{\cal Z}$.

 
 \vspace*{-0.35cm}
 
\section{Singularities of  Isobaric Partition  and Phase Transitions}

\vspace*{-0.2cm}

The IP, Eq.~(\ref{four}),  has two types of singularities:
1) the simple pole singularity
defined by the equation
\begin{equation}\label{pole}
s_g(T,\mu)~=~ {\cal F}(s_g,T,\mu)~,
\end{equation}
2)  the singularity  of the function ${\cal F}(s,T,\mu)$ 
 it-self at the point $s_l$ where the coefficient 
in linear over $k$ terms in the exponent is equal to zero,
\begin{equation}\label{sl}
s_l(T,\mu)~=~\frac{\tilde{\mu}}{Tb}~.
\end{equation}

The simple pole singularity corresponds to the gaseous phase 
where pressure is determined by the 
equation
\begin{eqnarray}\label{pgas}
p_g(T,\mu) &=& \left( \frac{mT }{2\pi}\right)^{3/2} T
\left[z_1 \exp\left(\frac{\mu-bp_g}{T}\right)
+ \sum_{k=2}^{\infty}
k^{ -\tau} \exp\left(
\frac{(\tilde{\mu} - bp_g)k - \sigma k^{2/3} }{T}\right)\right]~.
\end{eqnarray}
The singularity $s_l(T,\mu)$ of the function ${\cal F}(s,T,\mu)$
(\ref{Osix}) defines the liquid pressure
\begin{equation}\label{pl}
p_l(T,\mu)~\equiv~ T~s_l(T,\mu)~=~
\frac{\tilde{\mu}}{b}~.
\end{equation}

In the considered model the liquid phase is represented by an
infinite fragment, i.e. it corresponds to the macroscopic population
of the single mode $k = \infty$. Here one can see the analogy
with the Bose condensation where the  macroscopic population
of a single mode occurs in the momentum space.

In the $(T,\mu)$-regions where $\tilde{\mu} < bp_g(T,\mu)$ the gas phase
dominates ($p_g > p_l$), while  the liquid phase
corresponds to $\tilde{\mu} > b p_g(T,\mu)$. The liquid-gas phase transition
occurs when  two singularities coincide,
i.e. $s_g(T,\mu)=s_l(T,\mu)$.
A schematic  view of singular points is shown 
in Fig.~1a for $T <T_c$, i.e. when $\sigma > 0$.
The two-phase coexistence region is therefore defined by the
equation
\begin{equation}\label{ptr}
p_l(T,\mu)~=~p_g(T,\mu)~,~~~~{\rm i.e.,}~~ \tilde{\mu}~=~b~p_g(T,\mu)~.
\end{equation} 
One can easily see that ${\cal F}(s,T,\mu)$ is monotonously decreasing
function of $s$. 
The necessary condition for the phase
transition is that this function
remains finite in its singular
point \mbox{$s_l=\tilde{\mu}/Tb$:}
\begin{equation}\label{Fss}
{\cal F}(s_l,T,\mu)~<~\infty~.
\end{equation}
The convergence of ${\cal F}$ is determined
by $\tau$ and $\sigma$.
At $\tau=0$ the condition (\ref{Fss}) requires $\sigma(T) >0$.
Otherwise, ${\cal F}(s_l,T,\mu)=\infty$ and the simple pole
singularity $s_g(T,\mu)$ (\ref{pole}) is always the
rightmost $s$-singularity
of $\hat{\cal Z}$ (\ref{four}) (see Fig.~1b). 
At $T>T_c$, where $\sigma(T)|_{SMM}=0$, the considered system
can exist only in the one-phase state. It will be shown below
that for $\tau>1 $  the condition (\ref{Fss})
can be satisfied even at $\sigma(T)=0$.

At $T<T_c$ the system undergoes the 1-st order phase transition
across the line $\mu^*=\mu^*(T)$ defined by Eq.(\ref{ptr}).
Its explicit form is given
by the expression:
\begin{eqnarray}\label{muc}
\mu^*(T)~ & = &~ -~ W (T) ~
+~\left(\frac{mT}{2\pi}\right)^{3/2} T b
\left[z_1\exp\left(-~\frac{W(T)}{T}\right) 
+  \sum_{k=2}^{\infty}
k^{ -\tau} \exp\left(-~ \frac{\sigma~ k^{2/3}}{T}\right)
\right]~.
\end{eqnarray}
The points on the line
$\mu^*(T)$ correspond to the mixed phase
states. First we  consider the case  $\tau=- 1.5$ because it is  the standard SMM choice.

The
baryonic density 
is found as $(\partial p/\partial \mu)_T$ and
is given by the following  formulae in the liquid and gas phases
\begin{eqnarray}
& & \rho_l~  \equiv  ~
\left(\frac{\partial  p_l}{\partial \mu}\right)_{T}~
= ~ \frac{1}{b}~, \quad  \quad 
\rho_g  \equiv  ~
\left(\frac{\partial  p_g}{\partial \mu}\right)_{T}~=~
 \frac{ \rho_{id} }{ 1 + b\, \rho_{id} } ~,\label{rhog}
\end{eqnarray}
respectively. Here the function $ \rho_{id}$ is defined as
\begin{eqnarray}\label{rhoid}
\rho_{id}(T,\mu) & = & \left( \frac{mT }{2\pi}\right)^{3/2} 
\left[z_1 \exp\left(\frac{\mu-bp_g}{T}\right) 
 + \sum_{k=2}^{\infty} 
k^{1 -\tau} \exp\left(
\frac{( \tilde\mu - bp_g)k - \sigma k^{2/3}}{T} \right)\right]~.
\end{eqnarray}


\noindent
Due to the condition
(\ref{ptr})
this expression 
is simplified 
in the mixed phase: 
\begin{eqnarray}\label{rhoidmix}
\rho_{id}^{mix}(T)~ &\equiv&~\rho_{id}(T,\mu^*(T))~
=  \left( \frac{ mT }{2 \pi}\right)^{3/2}
 \left[z_1 \exp\left(-~\frac{W (T) }{T}\right)
~+~\sum_{k=2}^{\infty}
k^{1 - \tau}
\exp\left(-~ \frac{\sigma ~k^{2/3}}{T}\right)\right]~.
\end{eqnarray}
This formula clearly shows that the bulk (free) energy acts in favor of the composite
fragments, but the surface term favors single nucleons.

Since at $\sigma >0$ the sum in Eq.~(\ref{rhoidmix})
converges at any $\tau$, $\rho_{id}$ is finite and according to
Eq.~(\ref{rhog}) $\rho_g<1/b$. Therefore,
the baryonic density has a discontinuity $\Delta\rho =\rho_l-\rho_g >0$
across the line $\mu^*(T)$ (\ref{muc}) for  any $\tau$. 
The discontinuities
take place also for
the energy and entropy densities. 
The phase diagram of the system in the $(T,\rho)$-plane 
is shown in the upper panel of Fig.~2. 
The line
$\mu^*(T)$ (\ref{muc}) corresponding to the mixed phase
states 
is transformed into
the finite region in the $(T,\rho)$-plane. As usual,  in this mixed phase
region  of the
{ phase diagram the baryonic density $\rho$ and the energy density
are   superpositions of the corresponding densities of  liquid and gas:}
\begin{equation}\label{mixed}
\rho~=~\lambda~\rho_l~+~(1-\lambda)~\rho_g~, \quad \quad  
\varepsilon ~=~\lambda~\varepsilon_l~+~(1-\lambda)~\varepsilon_g~.
\end{equation}
Here $\lambda$ ($0<\lambda <1$) is a fraction of the system volume
occupied by the liquid  inside the mixed phase, 
{ and the  partial 
energy densities 
for $(i=l,g)$ can be found from the thermodynamic identity \cite{Bugaev:00}:}
\begin{equation}\label{eps}
\varepsilon_i~\equiv~T\frac{\partial p_i}{\partial T}~+~
\mu\frac{\partial p_i}{\partial \mu}~-~p_i~.
\end{equation}

Inside the mixed phase at constant density $\rho$ the
parameter $\lambda$ has a specific temperature dependence
shown in the upper panel of  Fig.~ 3:
from an approximately
constant value $\rho/\rho_{\rm{o}}$ at small $T$ the function 
$\lambda(T)$ drops to zero in a narrow
vicinity of the boundary separating the  mixed phase and 
the pure gaseous phase.
This corresponds to a fast change of the configurations from
the state which is  dominated by one infinite liquid fragment to 
the gaseous multifragment configurations. It happens inside the
mixed phase  without
discontinuities in the thermodynamical functions.

An abrupt decrease of $\lambda(T)$ near this boundary
causes a strong
increase of the energy density as a function of temperature.
This is evident from the middle panel of  Fig.~3 which shows the caloric curves at different
baryonic densities. One can clearly see a 
leveling of temperature at energies per nucleon between 10 
and 20 MeV.
As a consequence this  
leads to a sharp peak 
in the specific heat per nucleon at constant density,
$c_{\rho}(T)\equiv (\partial \varepsilon/\partial T)_{\rho}/\rho~$,
presented in Fig. 3.
A finite discontinuity of $c_{\rho}(T)$ arises
at the boundary between the mixed phase and  the gaseous phase.
This finite discontinuity
is caused by the fact that
$\lambda(T)=0$, but
$(\partial\lambda/\partial T)_{\rho} \neq 0$
at this boundary 
(see Fig. 3).

It should be emphasized that the energy density is continuous
at the boundary of the mixed phase and the gaseous phase, hence
the sharpness of the
peak in $c_{\rho}$ is entirely due to the strong temperature
dependence
of $\lambda(T)$ near this boundary. 
Moreover, at any $\rho < \rho_{\rm o}$
the maximum value of $c_{\rho}$ remains finite
and the peak width in $c_{\rho}(T)$ is nonzero in the thermodynamic
limit considered in our study. 
This is in contradiction with the expectation of Refs. \cite{Gupta:98,Gupta:99}
that an infinite peak of zero width will appear in $c_{\rho}(T)$ in this
limit.
Also a comment about the so-called ``boiling point''
is appropriate here.
This is a discontinuity in the energy density $\varepsilon(T)$ 
or, equivalently, a plateau in the
temperature as a function of the excitation energy. 
Our analysis shows that this type of behavior indeed happens 
at constant pressure, but not at constant density! This is similar to
the usual picture of a liquid-gas phase transition.
In Refs. \cite{Gupta:98,Gupta:99} a rapid  
increase of the energy density as a function of temperature
at fixed $\rho$ near the boundary of the mixed and gaseous phases
(see the lower panel of  Fig.~3)
was misinterpreted as a manifestation of the ``boiling point''.

New possibilities appear
at non-zero values of the parameter $\tau$.  
At $0<\tau \le 1$ the qualitative picture remains the same
as discussed above, although there are some
quantitative changes. 
For $\tau > 1$  the condition (\ref{Fss}) is also satisfied 
at $T>T_c$ where $\sigma(T)|_{SMM} =0$.
Therefore, the liquid-gas phase transition
extends now to all temperatures. Its properties
are, however, different for $\tau >  2 $ and for $\tau \le  2$
(see Fig.~2). 
If $\tau > 2$ the 
gas density is always lower than $1/b$ as $\rho_{id}$ is finite.
Therefore, the
liquid-gas transition at $T>T_c$ remains
the 1-st order phase transition with discontinuities
of baryonic density, entropy and energy densities (lower panel in Fig.~2) .
%



\section{The Critical Indices  and Scaling Relations of the SMM}


The above results allow one to find 
the critical exponents $\alpha^\prime, \, \beta$ and $\gamma^\prime$ of the simplified SMM. 
These exponents  describe the temperature dependence of the system near the critical point on the coexistence curve $\mu^* = \mu^*(T)$  (\ref{ptr}), where the effective chemical potential vanishes  
$\nu \equiv  \mu^*(T) + W(T) -  b p(T,  \mu^*(T) )  = 0 $
\begin{eqnarray}\label{alpha}
c_\rho &~ \sim ~& 
\left\{
\begin{tabular}{ll}
$\mid \varepsilon \mid^{-\alpha}$\,, \hspace*{0.975cm} ${\rm for} ~ \varepsilon  < 0 $~, \\
$\varepsilon^{-\alpha^\prime}$\, ,\hspace*{1.25cm} ${\rm for} ~ \varepsilon  \ge 0$~,
\end{tabular}
\right.
\\ \label{beta}
\Delta\rho &~ \sim ~ & 
\varepsilon^{\beta}\,,   \hspace*{2.15cm}	 {\rm for} ~ \varepsilon  \ge 0 ~,
\\ \label{gamma}
\kappa_T &~ \sim ~&  
\varepsilon^{-\gamma^\prime}\,,\hspace*{1.925cm} 	 {\rm for} ~  \varepsilon  \ge 0 ~,
\end{eqnarray}
where $\Delta\rho \equiv \rho_l - \rho_g$ defines the order parameter, $c_\rho \equiv \frac{T}{\rho} \left(\frac{\partial s}{\partial T} \right)_\rho$ denotes the specific heat at fixed particle density and $\kappa_T \equiv \frac{1}{\rho} \left(\frac{\partial \rho}{\partial p} \right)_T$ is the  isothermal compressibility.
The shape of the critical isotherm for $\rho \leq \rho_c$ is given by the critical index $\delta$
(the tilde indicates $\varepsilon = 0$ hereafter)
\begin{equation}\label{delta}
p_c - \tilde{p}  ~\sim ~
(\rho_c  - \tilde{\rho})^{\delta}		\hspace*{1.1cm} {\rm for} ~  \varepsilon = 0~. 
\end{equation}

The calculation of $\alpha$ and $\alpha^\prime$ requires the specific heat $c_{\rho}$. With the formula \cite{Ya:64}
\begin{equation}\label{crho}
\frac{c_{\rho}(T, \mu)}{T}~= ~ \frac{1}{\rho}\left(\frac{\partial^2 p}{\partial T^2} \right)_{\rho} - \left(\frac{\partial^2 \mu}{\partial T^2} \right)_{\rho}
\end{equation}
one obtains the specific heat on the PT curve by replacing the partial derivatives by the total ones \cite{Fi:70}. The latter can be done for every state inside or on the boundary of the mixed phase region.
For the chemical potential $\mu^*(T) = bp^*(T) - W(T)$ one gets
%
$
\frac{c_{\rho}^*(T)}{T}~ = ~ 
\left( \frac{1}{\rho}- b \right) \frac{ {\rm d}^2 p^*(T)}{ {\rm d} T^2} + \frac{{\rm d}^2 W(T)}{{\rm d} T^2}.
$
%
Here the asterisk indicates the condensation line ($\nu = 0$) hereafter.  
Fixing $\rho = \rho_c = \rho_l = 1/b$ one finds $c_{\rho_l}^*(T) = T\frac{ {\rm d^2}W(T)} {{\rm d}T^2}$ and, hence, obtains 
$ \alpha ~= ~\alpha^\prime ~= ~ 0.$
To calculate $\beta$, $\gamma^\prime$  and $\delta$ the behavior  
of the series 
\begin{equation}\label{I}
\mys{q}( \varepsilon, \nu)  \equiv 
\sum_{k=2}^{\infty}~k^{q-\tau}~e^{\textstyle \frac{\nu}{T_c} k - A \varepsilon^\zeta k^\sigma}
\end{equation}
should be analyzed 
for  small positive values of $\varepsilon$  and $- \nu$ \mbox{$(A \equiv a_{\rm o}/T_c)$}.
In the limit $\varepsilon \rightarrow 0$ the function $\mys{q}( \varepsilon, 0)$ remains finite, if $\tau > q+1$, and diverges otherwise. For $\tau = q+1$ this divergence is logarithmic. 
{ The case $\tau < q + 1$ is analyzed in some details, since even in Fisher's papers it was 
performed incorrectly.}

With the substitution ${\mbox z_{k}\equiv k\left[ A \varepsilon^\zeta \right]^{1/\sigma} }$
{ one can prove \cite{Reuter:01}   that in the limit  $\varepsilon \rightarrow 0$ 
the series on the r.\,h.\,s. of (\ref{I}) converges to an integral}

\vspace*{-0.5cm}

\begin{equation}\label{series}
\hspace*{1.4cm}
\mys{q}( \varepsilon, 0)~= ~\left[ A  \varepsilon^\zeta \right]^{\textstyle \frac{\tau - q }{\sigma}}~ 
\sum_{k=2}^{\infty}~
z_k^{q-\tau}~e^{\textstyle -z_k^\sigma}~ \rightarrow ~
\left[ A  \varepsilon^\zeta \right]^{\textstyle \frac{\tau-q-1}{\sigma}}
\hspace*{-3 mm}\int \limits_{2 \left[ A  \varepsilon^\zeta \right]^{\frac{ 1}{\sigma}}}^{\infty}
\hspace*{-3 mm} {\rm d}z~
z^{q-\tau}~e^{\textstyle -z^\sigma} ~.
\end{equation}

\vspace*{-0.1cm}

\noindent
The assumption $q-\tau > -1$ is { sufficient} to guarantee  the convergence of the integral at its lower limit.
{ Using this representation, one finds the following general results}  \cite{Reuter:01}
\begin{equation}\label{Prop1}
\mys{q}( \varepsilon, 0) \sim \left\{ 
\begin{array}{ll}
\varepsilon^{\textstyle\, \frac{\zeta}{ \sigma }(\tau - q- 1) }\,,
&
{\rm if} ~ \tau < q + 1 \,,  \\
\ln\mid\varepsilon\mid\,,  
&
{\rm if} ~ \tau = q + 1 \,,
\\
\varepsilon^{\,0}\,, 
&
{\rm if} ~  \tau > q + 1 \, .
\end{array}
\right. \quad {\rm and} \quad 
\mys{q}( 0, \tilde{\nu})~\sim ~\left\{ 
\begin{array}{ll}
\tilde{\nu}^{\textstyle\, \tau - q- 1}\,,
&
{\rm if} ~ \tau < q +1\,,  
\\
\ln\mid\tilde{\nu}\mid\,,  
&
{\rm if} ~ \tau = q + 1 \,,
\\
\tilde{\nu}^{\,0}\,,
&
{\rm if} ~  \tau > q + 1 \,,
\end{array}
\right.
\end{equation}
which allowed us to find out the critical indices of the SMM (see  Table 1).


\noindent{\footnotesize{\bf
T a b l e~ 1.
Critical exponents of the SMM and FDM as functions of Fisher index  \boldmath$\tau$
for  the general parametrization of the  surface energy 
 \boldmath$ \sigma(T) k^{\frac{2}{3} }  \rightarrow  \varepsilon^\zeta k^\sigma $  with  \boldmath$\varepsilon = (T_c - T)/T_c$
} \tabcolsep=17.5pt

\noindent
\begin{tabular}{cccccc}\hline
& \multicolumn{1}{c}{ $\alpha^\prime$ } &
   \multicolumn{1}{c}{ $\alpha^\prime_s$ } &
\multicolumn{1}{c}{ $\beta$ } &
\multicolumn{1}{c}{ $\gamma^\prime$ } &
\multicolumn{1}{c}{ $\delta$ } \\
\hline
SMM for  $\tau < 1 + \sigma$ & $0$ &  $2 -  \frac{\zeta}{\sigma}$     &  $ \frac{\zeta}{\sigma}  (2 - \tau) $    
& $ \frac{2 \zeta}{\sigma}  \left(\tau - \frac{3}{2} \right) $   &  $\frac{\tau - 1}{2 - \tau}$   \\ 
SMM for  $\tau \ge 1 + \sigma$ & $0$ &  $2 -  \frac{\zeta}{\sigma}(\sigma + 2 - \tau) $     &  $ \frac{\zeta}{\sigma}  (2 - \tau) $    
& $ \frac{2 \zeta}{\sigma}  \left(\tau - \frac{3}{2} \right) $   &  $\frac{\tau - 1}{2 - \tau}$   \\ 
FDM  &  $2 -  \frac{\zeta}{\sigma}{ (\tau - 1)}$ & N/A & $ \frac{\zeta}{\sigma}  (\tau -2 ) $  & $ \frac{\zeta}{\sigma}  (3 - \tau) $ &     $\frac{1}{\tau - 2 }$   \\
\hline
\end{tabular}
\label{tab:a}
}
\vskip8mm

%

In the special case $\zeta = 2\sigma$
 the well-known exponent inequalities proven for real gases by
\begin{eqnarray}
\label{F1}
{\rm Fisher  [36]:} & \hspace{.5cm}	\alpha^\prime + 2\beta + \gamma^\prime ~& \ge~2 ~,\\
\label{G}
{\rm Griffiths [37]:}& 		\alpha^\prime + \beta(1 + \delta) & \ge~ 2 ~,\\
\label{L}
{\rm Liberman [38]:}&			\gamma^\prime + \beta(1-\delta) & \ge ~0 ~,
\end{eqnarray}
are fulfilled exactly for any $\tau$. (The corresponding exponent inequalities for magnetic systems are often called Rushbrooke's, Griffiths' and Widom's inequalities, respectively.) For $\zeta > 2\sigma $,  Fisher's and Griffiths' exponent inequalities are fulfilled as inequalities and for $\zeta < 2\sigma $ they are not fulfilled. The contradiction to Fisher's and Griffiths' exponent inequalities in this last case is not surprising. This is due to the fact that in the present version of the SMM the critical isochore $\rho = \rho_c = \rho_l$ lies on the boundary of the mixed phase to the liquid. Therefore, in  expression (2.13) in Ref. \cite{Fi:64} for the specific heat only the liquid phase contributes and, 
{ therefore, Fisher's 
 proof of Ref. \cite{Fi:64} 
 } following (2.13) cannot be applied for the SMM. Thus, the exponent inequalities (\ref{F1}) and (\ref{G}) have to be modified for the SMM. Using  results of  Table \ref{tab:a},  
one finds the following scaling relations
\begin{eqnarray}
\alpha^\prime + 2\beta + \gamma^\prime  =   \frac{\zeta}{\sigma} 
\hspace{0.5cm} {\rm and} \hspace{0.5cm}
\alpha^\prime + \beta(1 + \delta)  =  \frac{\zeta}{\sigma} ~.
\end{eqnarray}
Liberman's exponent inequality (\ref{L}) is fulfilled exactly for any choice of $\zeta$ and $\sigma$.

Since the coexistence curve of the SMM is not symmetric with respect to $\rho = \rho_c$, it is interesting with regard to the specific heat to consider the difference $\Delta c_\rho(T) \equiv c^*_{\rho_g}(T) - c^*_{\rho_l}(T)$, following the suggestion of Ref. \cite{Fi:70}. 
Using  Eq. (\ref{crho}) for gas and liquid  and noting that $1/\rho_g^* - b = 1/\rho_{id}^*$,  one obtains a specially defined index  $\alpha^\prime_s$
from the most divergent term for $\zeta>1$
\begin{eqnarray}\label{alphas}
\Delta c_\rho(T) ~=~ \frac{T}{\rho_{id}^*(T)} \frac{ {\rm d}^2 p^*(T)}{ {\rm d} T^2}~ \quad \Rightarrow
\quad  \alpha^\prime_s ~= ~
\left\{ 
\begin{array}{ll}
\vspace{0.1cm}2 - \frac{\zeta}{\sigma}\,,
& 
{\rm if} ~ \tau < \sigma + 1 \,,  \\
2 - \frac{\zeta}{\sigma}( \sigma + 2 - \tau)\,,	
&
{\rm if} ~  \tau \ge \sigma + 1 \,.
\end{array}
\right.
\end{eqnarray}
Then it is $\alpha_s^\prime > 0$ for $\zeta/\sigma <2$.
Thus, approaching the critical point along any isochore within the mixed phase region except for $\rho = \rho_c = 1/b$ the specific heat diverges for $\zeta/\sigma <2$ as defined by $\alpha^\prime_s$ and remains finite for the isochore $\rho = \rho_c = 1/b$. This demonstrates the exceptional character of the critical isochore in this model. 

In the special case that $\zeta = 1$ one finds $\alpha^\prime_s = 2 - 1/\sigma$ for $\tau \leq 1 + 2 \sigma$ and $\alpha^\prime_s = -\beta$ for $\tau > 1 + 2 \sigma$.
Therefore, using $\alpha_s^\prime$ instead of $\alpha^\prime$, the exponent inequalities (\ref{F1}) and (\ref{G}) are fulfilled
exactly if $\zeta >1$ and  $\tau \leq \sigma + 1$ or if $\zeta =1$ and  $\tau \leq 2\sigma + 1$. In all other cases (\ref{F1}) and (\ref{G}) are fulfilled as inequalities.
Moreover, it can be shown that the SMM belongs to the universality class of real gases for $\zeta >1$ and $\tau \ge \sigma + 1$.

The comparison of 
the above derived formulae for the critical exponents of the SMM
for $\zeta = 1$
with  
those obtained within the FDM 
(Eqs. 51-56 in \cite{Fisher:67}) shows that these models
belong to  different  universality classes (except for the singular case $\tau = 2$). 

Furthermore, one has to note that for $\zeta = 1\,,~\sigma \leq 1/2$ and $1 + \sigma < \tau \leq 1 + 2 \sigma$ the critical exponents of the SMM coincide with those of the exactly solved one-dimensional FDM with non-zero droplet-volumes \cite{Fi:70}. 

For the usual parameterization of the SMM \cite{Bondorf:95} 
one obtains with $\zeta = 5/4$ and $\sigma = 2/3$ the exponents
\begin{eqnarray}
\alpha^\prime_s ~&=& ~\left\{ 
\begin{array}{ll}
\vspace{0.2cm} \frac{1}{8}\,, 
&
{\rm if} ~ \tau < \frac{5}{3}   \\
\frac{15}{8}\tau - 3\,, 
&
{\rm if} ~  \tau \ge \frac{5}{3}  
\end{array}
\right.~,\hspace*{0.5cm}
\beta~=~
\frac{15}{8}\left(2 - \tau \right)~, \hspace*{0.5cm}
\gamma^\prime ~ = ~
\frac{15}{4}\left( \tau - \frac{3}{2} \right)~,\hspace*{0.5cm}
\delta ~ = ~ \frac{\tau - 1}{2 - \tau}~. 
\end{eqnarray}
{Thus, Fisher suggestion to use $\alpha^\prime_s$ instead of  $\alpha^\prime$
allows one to  {\it ``save''  the exponential inequalities}, however,  it is not a final  solution of the problem.
 }

The critical indices of the nuclear liquid-gas PT were determined 
from the multifragmentation of gold nuclei \cite{Eos:94} and
found to be close to those ones of real gases.  
The method used to extract the critical exponents 
$\beta$ and $\gamma^\prime$  in Ref. \cite{Eos:94} was, however,
found to have large uncertainties of about 25 per cents \cite{Bau:94}.
Nevertheless, those results allow us  
to estimate the value of $\tau$
from the experimental values of the critical exponents  of real
gases taken with large error bars.
Using the above results we 
generalized \cite{Reuter:01}  the exponent relations of Ref. \cite{Fi:70}
\begin{eqnarray}
\label{ntau1}
\tau ~= ~ 2 - \frac{\beta}{ \gamma^\prime + 2 \beta} ~~~ {\rm and} \hspace{0.5cm} 
\tau ~= ~ 2 - \frac{1}{1 + \delta}
\end{eqnarray}
for arbitrary $\sigma$ and $\zeta$. 
Then, one obtains with \cite{Hu} $\beta = 0.32-0.39$\,, $\gamma^\prime = 1.3-1.4$ and $\delta = 4-5$ the estimate $\tau =  1.799 - 1.846$. 
This demonstrates also that the value of $\tau$ is rather 
insensitive to the special choice of $\beta\,,~\gamma^\prime$ and $\delta$, which leads to  $\alpha^\prime_s \cong 0.373 - 0.461$ for the SMM.
Theoretical values for $\beta\,,~\gamma^\prime$ and $\delta$ for Ising-like systems within the renormalized $\phi^4$ theory \cite{Zi:98} lead to the narrow range $\tau = 1.828 \pm 0.001$\,.
The values of  $\beta\,,~\gamma^\prime$ and $\delta$ indices for nuclear matter
and percolation of  two- and three-dimensional clusters are reviewed in \cite{Elliott:05wci}.

There was a decent  try to study 
the critical indices of  the SMM  numerically 
\cite{El:00}. The version V2 of Ref. \cite{El:00} corresponds  
precisely to 
our model with $\tau = 0$, $\zeta = 5/4$ and $\sigma = 2/3$,
but their results contradict to our analysis.
Their results for version V3 of Ref. \cite{El:00} are in contradiction with
our proof presented in Ref. \cite{Bugaev:00}. There it was shown that for non-vanishing surface energy (as in version V3)
the critical point does not exist at all.
The latter was found in \cite{El:00}
for the finite system
and the critical indices were analyzed.
Such a strange result, on one hand,  
 indicates 
that the numerical  methods used in Ref. \cite{El:00} 
are not  self-consistent, and, on the other hand, it shows an indispensable value of the analytical
calculations, which can be used as a test problem  for numerical algorithms.

It is widely believed that the effective value of $\tau$ defined 
as $\tau_{\rm eff} \equiv  - \partial \ln n_k(\varepsilon) / \partial \ln k$ 
 attains
its minimum at the critical point (see references in \cite{EOS:00}).
This has been shown for the version of the FDM with the constraint of sufficiently small surface tension $a \cong 0$ for $T\ge T_c$ \cite{Pan:83} and also can be seen easily for the SMM. Taking the SMM fragment distribution  
$n_k(\varepsilon) = g(T) k^{-\tau} \exp[{\textstyle \frac{\nu}{T} k 
- \frac{a(\varepsilon)}{T} k^\sigma }] \sim k^{- \textstyle \tau_{\rm eff}}  $ 
one finds
\begin{equation}\label{teff}
\tau_{\rm eff}~ = ~ \tau - \frac{\nu}{T} k + \frac{\sigma a(\varepsilon)}{T} k^\sigma  
\quad \Rightarrow \quad \tau = \min( \tau_{\rm eff}) ~,
\end{equation}
where the last step follows from the fact that  
the inequalities 
$a(\varepsilon) \ge 0$\,, $\nu \leq 0$ become equalities  
at the critical point $\nu = a(0) = 0$. 
Therefore,
the SMM predicts
that the minimal value of $\tau_{\rm eff}$ corresponds to the critical point,
where, in contrast to other regions of parameters,  
the mass distribution of fragments should become power-like.

In the E900 $\pi^-+$\,Au multifragmentation experiment \cite{ISIS} 
the ISiS collaboration measured
the dependence of
$\tau_{\rm eff}$ upon the excitation energy and 
found the minimum value ${\rm min}(\tau_{\rm eff} ) \cong 1.9$ 
(Fig.\,5 of Ref. \cite{ISIS} ). 
Also the EOS collaboration \cite{EOS:00} performed an analysis of the minimum of $\tau_{\rm eff}$ on Au\,+\,C multifragmentation data. The fitted $\tau_{\rm eff}$, plotted in Fig.\,11.b of Ref. \cite{EOS:00} versus the fragment multiplicity, exhibits a minimum in the range 
${\rm min}(\tau_{\rm eff}) \cong 1.8 - 1.9$\,. 
Both results contradict the original FDM \cite{Fisher:67}, but agree well
with the above estimate of $\tau$ for real gases and for Ising-like systems in general.


\vspace*{-0.35cm}

\section{Constrained SMM in Finite Volumes}

\vspace*{-0.25cm}
{
Despite the  great
success, the application of the exact  solution \cite{Bugaev:00,Bugaev:01,Reuter:01}
to the description of experimental data is  limited because this solution  
corresponds to an infinite system and due to that it cannot  account for a more
complicated interaction between nuclear fragments. 
Therefore, it was necessary to extend  the exact  solution \cite{Bugaev:00,Bugaev:01,Reuter:01}
to finite volumes.  It is clear that for the finite volume extension it is necessary 
to account for the finite size  and geometrical shape of 
the largest fragments,  when  they  are comparable with the system volume.  
For this one has to abandon the arbitrary size of largest fragment and
consider the constrained SMM (CSMM) in which
the largest fragment  size
is explicitly related to the volume $V$ of the system. 
}
Thus, the CSMM  assumes  a more strict constraint
$\sum\limits_k^{K(V)} k~n_k =A$ , where the size
of the largest fragment  $K(V) = \alpha V/b$ cannot exceed the total volume of the system 
(the parameter $\alpha \le 1$  is introduced for convenience).
 The case of the fixed size of the largest fragment, i.e.  $K(V) = Const$,   analyzed  numerically 
  in Ref. \cite{CSMM} 
 is also included in our treatment. 
A similar restriction should be also applied to the upper limit of the product in 
all partitions $Z_A^{id} (V,T)$, $Z_A(V,T)$
and ${\cal Z}(V,T,\mu)$ introduced above 
(how to deal with the real values of $K(V)$, see later). 
Then the  model with this constraint, the CSMM,  cannot be solved by the Laplace 
transform method, because the volume integrals cannot be evaluated due to a complicated 
functional $V$-dependence.  
However, the CSMM can be solved analytically with the help of  the following identity  \cite{Bugaev:04a}
\vspace*{-0.2cm}
\begin{equation}\label{nfour}
G (V) = 
%
\int\limits_{-\infty}^{+\infty} d \xi~ \int\limits_{-\infty}^{+\infty}
  \frac{d \eta}{{2 \pi}} ~ 
{\textstyle e^{ i \eta (V - \xi) } } ~ G(\xi)\,, 
\end{equation}
\vspace*{-0.3cm}

\noindent
which is based on the Fourier representation of the Dirac $\delta$-function. 
The representation (\ref{nfour}) allows us to decouple the additional volume
dependence and reduce it to the exponential one,
which can be dealt by the usual Laplace transformation
in the  following sequence of steps
\vspace*{-0.3cm}
\begin{eqnarray} \label{nfive}
&&\hat{\cal Z}(\lambda,T,\mu)~\equiv ~\int_0^{\infty}dV~{\textstyle e^{-\lambda V}}
~{\cal Z}(V,T,\mu) = 
\hspace*{-0.0cm}\int_0^{\infty}\hspace*{-0.2cm}dV^{\prime}
\int\limits_{-\infty}^{+\infty} d \xi~ \int\limits_{-\infty}^{+\infty}
\frac{d \eta}{{2 \pi}} ~ { \textstyle e^{ i \eta (V^\prime - \xi) - \lambda V^{\prime} } } \times 
\nonumber \\
&& \sum_{\{n_k\}}\hspace*{-0.1cm} \left[\prod_{k=1}^{K( \xi)}~\frac{1}{n_k!}~\left\{V^{\prime}~
{\textstyle \phi_k (T) \,  
e^{\frac{ (\mu  - (\lambda - i\eta) bT )k}{T} }}\right\}^{n_k} \right] \Theta(V^\prime) = 
\hspace*{-0.0cm}\int_0^{\infty}\hspace*{-0.2cm}dV^{\prime}
\int\limits_{-\infty}^{+\infty} d \xi~ \int\limits_{-\infty}^{+\infty}
  \frac{d \eta}{{2 \pi}} ~ { \textstyle e^{ i \eta (V^\prime - \xi) - \lambda V^{\prime} 
+ V^\prime {\cal F}(\xi, \lambda - i \eta) } }\,.
%
%
%
\end{eqnarray}
\vspace*{-0.3cm}

\noindent
After changing the integration variable $V \rightarrow V^{\prime} = V - b \sum\limits_k^{K( \xi)} k~n_k $,
the constraint of $\Theta$-function has disappeared.
Then all $n_k$ were summed independently leading to the exponential function.
Now the integration over $V^{\prime}$ in Eq.~(\ref{nfive})
can be straightforwardly done resulting in
\vspace*{-0.2cm}
\begin{equation}\label{six}
\hspace*{-0.4cm}\hat{\cal Z}(\lambda,T,\mu) = \int\limits_{-\infty}^{+\infty} \hspace*{-0.1cm} d \xi
\int\limits_{-\infty}^{+\infty} \hspace*{-0.1cm}
\frac{d \eta}{{2 \pi}} ~ 
\frac{  \textstyle e^{ - i \eta \xi }  }{{\textstyle \lambda - i\eta ~-~{\cal F}(\xi,\lambda - i\eta)}}~,
\end{equation}

\vspace*{-0.3cm}

\noindent
where the function ${\cal F}(\xi,\tilde\lambda)$ is defined as follows 
\vspace*{-0.2cm}
\begin{eqnarray}\label{seven}
&&\hspace*{-0.4cm}{\cal F}(\xi,\tilde\lambda) = \sum\limits_{k=1}^{K(\xi) } \phi_k (T) 
~e^{\frac{(\mu  - \tilde\lambda bT)k}{T} }
= 
\hspace*{-0.0cm}\left(\frac{m T }{2 \pi} \right)^{\frac{3}{2} } \hspace*{-0.1cm} \biggl[ z_1
~{\textstyle e^{ \frac{\mu- \tilde\lambda bT}{T} } } + \hspace*{-0.1cm} \sum_{k=2}^{K(\xi) }
k^{-\tau} e^{ \frac{(\mu + W - \tilde\lambda bT)k - \sigma k^{2/3}}{T} }  \biggr]\,.\,
\end{eqnarray}

As usual, in order to find the GCP by  the inverse Laplace transformation,
it is necessary to study the structure of singularities of the isobaric partition (\ref{seven}).


\vspace*{-0.35cm}

\section{Isobaric Partition Singularities at Finite Volumes}

\vspace*{-0.2cm}

The isobaric partition (\ref{seven}) of the CSMM is, of course, more complicated
than its SMM analog \cite{Bugaev:00,Bugaev:01}
because for finite volumes the structure of singularities in the CSMM 
is much richer than in the SMM, and they match in the limit $V \rightarrow \infty$ only.
To see this let us first make the inverse Laplace transform:
\vspace*{-0.2cm}
\begin{eqnarray}\label{eight}
\hspace*{-0.0cm}{\cal Z}(V,T,\mu)~ = 
\int\limits_{\chi - i\infty}^{\chi + i\infty}
\frac{ d \lambda}{2 \pi i} ~ \hat{\cal Z}(\lambda,  T, \mu)~ e^{\textstyle   \lambda \, V } & = &
\hspace*{-0.0cm}\int\limits_{-\infty}^{+\infty} \hspace*{-0.1cm} d \xi
\int\limits_{-\infty}^{+\infty} \hspace*{-0.1cm}  \frac{d \eta}{{2 \pi}}  
\hspace*{-0.1cm} \int\limits_{\chi - i\infty}^{\chi + i\infty}
\hspace*{-0.1cm} \frac{ d \lambda}{2 \pi i}~ 
\frac{\textstyle e^{ \lambda \, V - i \eta \xi } }{{\textstyle \lambda - i\eta ~-~{\cal F}(\xi,\lambda - i\eta)}}~= 
\nonumber \\
&&\hspace*{-0.0cm}\int\limits_{-\infty}^{+\infty} \hspace*{-0.1cm} d \xi
\int\limits_{-\infty}^{+\infty} \hspace*{-0.1cm}  \frac{d \eta}{{2 \pi}}
\,{\textstyle e^{  i \eta (V - \xi)  } } \hspace*{-0.1cm} \sum_{\{\lambda _n\}}
e^{\textstyle  \lambda _n\, V } 
{\textstyle 
\left[1 - \frac{\partial {\cal F}(\xi,\lambda _n)}{\partial \lambda _n} \right]^{-1} } \,,
\end{eqnarray}
\vspace*{-0.3cm}

noindent
where the contour  $\lambda$-integral is reduced to the sum over the residues of all singular points
$ \lambda = \lambda _n + i \eta$ with $n = 1, 2,..$, since this  contour in the complex $\lambda$-plane  obeys the
inequality $\chi > \max(Re \{  \lambda _n \})$.  
Now both remaining integrations in (\ref{eight}) can be done, and the GCP becomes 
\vspace*{-0.2cm}
\begin{equation}\label{nine}
{\cal Z}(V,T,\mu)~ = \sum_{\{\lambda _n\}}
e^{\textstyle  \lambda _n\, V }
{\textstyle \left[1 - \frac{\partial {\cal F}(V,\lambda _n)}{\partial \lambda _n} \right]^{-1} } \,,
\end{equation}
\vspace*{-0.2cm}

\noindent 
i.e. the double integral in (\ref{eight}) simply  reduces to the substitution   $\xi \rightarrow V$ in
the sum over singularities. 
This is a remarkable result which 
was  formulated in Ref. \cite{Bugaev:04a}  as the following 
\underline{\it theorem:}
{\it if the Laplace-Fourier image of the excluded volume GCP exists, then
for any additional $V$-dependence of ${\cal F}(V,\lambda _n)$ or $\phi_k(T)$
the GCP can be identically represented by Eq. (\ref{nine}).}


The simple poles in (\ref{eight}) are defined by the  equation 
\begin{equation}\label{ten}
\lambda _n~ = ~{\cal F}(V,\lambda _n)\,.
\end{equation}
In contrast to the usual SMM \cite{Bugaev:00,Bugaev:01} the singularities  $ \lambda _n $ 
are (i) 
 are volume dependent functions, if $K(V)$ is not constant,
and (ii) they can have a non-zero imaginary part, but 
in this case there  exist  pairs of complex conjugate roots of (\ref{ten}) because the GCP is real.

Introducing the real $R_n$ and imaginary $I_n$ parts of  $\lambda _n = R_n + i I_n$,
we can rewrite  Eq. (\ref{ten})
as a system of coupled transcendental equations 
\vspace*{-0.2cm}
\begin{eqnarray}\label{eleven}
&&\hspace*{-0.2cm} R_n = ~ \sum\limits_{k=1}^{K(V) } \tilde\phi_k (T)
~{\textstyle e^{\frac{Re( \nu_n)\,k}{T} } } \cos(I_n b k)\,,
\\
\label{twelve}
&&\hspace*{-0.2cm} I_n = - \sum\limits_{k=1}^{K(V) } \tilde\phi_k (T)
%
%
~{\textstyle e^{\frac{Re( \nu_n)\,k}{T} } } \sin(I_n b k)\,,
\end{eqnarray}
\vspace*{-0.3cm}

\noindent
where we have introduced the 
set of  the effective chemical potentials  $\nu_n  \equiv  \nu(\lambda_n ) $ with $ \nu(\lambda) = \mu + W (T)  - \lambda b\,T$, and 
the reduced distributions $\tilde\phi_1 (T) = \left(\frac{m T }{2 \pi} \right)^{\frac{3}{2} }  z_1 \exp(-W(T)/T)$ and 
$\tilde\phi_{k > 1} (T) = \left(\frac{m T }{2 \pi} \right)^{\frac{3}{2} }  k^{-\tau}\, \exp(-\sigma (T)~ k^{2/3}/T)$ for convenience.


Consider the real root $(R_0 > 0, I_0 = 0)$, first. 
For $I_n = I_0 = 0$ the real root $R_0$ exists for any $T$ and $\mu$.
Comparing $R_0$ with the expression for vapor pressure of the analytical SMM solution 
\cite{Bugaev:00,Bugaev:01}
shows   that $T R_0$ is  a constrained grand canonical pressure of the gas. 
 As usual,  for  finite volumes the total mechanical pressure \cite{Hill,Bugaev:04a} differs from   $T R_0$.
Equation (\ref{twelve}) shows that for $I_{n>0} \neq 0$ the inequality $\cos(I_n b k) \le 1$ never 
become the equality for all $k$-values  simultaneously. Then from Eq. (\ref{eleven})  
one obtains ($n>0$)
\begin{equation}\label{thirteen}
R_n < \sum\limits_{k=1}^{K(V) } \tilde\phi_k (T)
~{\textstyle e^{\frac{Re(\nu_n)\, k}{T} } } \quad \Rightarrow \quad R_n < R_0\,, 
\end{equation}
where the second inequality (\ref{thirteen}) immediately follows from the first one.
{ In other words, the gas singularity is always the rightmost one.
This fact }
plays a decisive role in the thermodynamic limit
$V \rightarrow \infty$.

The interpretation of the complex roots $\lambda _{n>0}$  is less straightforward.
According to Eq. (\ref{nine}),   the  GCP is a superposition of  the
states of different  free energies $- \lambda _n V T$.  
(Strictly speaking,  $- \lambda _n V T$  has  a meaning of  the change of free energy, but
we  will use the  traditional  term for it.)
For $n>0$ the free energies  are complex. 
Therefore,  
 $-\lambda _{n>0} T$ is   the density of free energy.  The real part  of the free energy density,
  $- R_n T$, defines the significance
 of the state's  contribution to the partition:  due to (\ref{thirteen}) 
 the  largest contribution  always comes from the gaseous state and
 has the smallest  real part  of free energy density. As usual,  the states which do not have
 the smallest value of the (real part of)  free energy, i. e.  $- R_{n>0} T$, are thermodynamically metastable. 
 For  infinite   volume 
 they should not contribute unless they are infinitesimally close to  $- R_{0} T$, 
 but for finite volumes their contribution to the GCP may be important.

As one sees from (\ref{eleven}) and (\ref{twelve}), the states of  different  free energies have  
different values of the effective chemical potential $\nu_n$, which is not the case for
infinite volume \cite{Bugaev:00,Bugaev:01},
where there  exists a single value for the effective chemical potential. 
Thus,  for finite $V$
the  states which contribute to the GCP (\ref{nine}) are not in a true chemical equilibrium.

{The meaning of the imaginary part of the free energy density  becomes  clear from 
(\ref{eleven}) and (\ref{twelve}) \cite{Bugaev:05csmm}: as one can see from (\ref{eleven})  
the imaginary part $I_{n>0}$
effectively changes the number of degrees of freedom of  each $k$-nucleon fragment ($k \le K(V)$)
contribution to  the free energy  density  $- R_{n>0} T$.  It is clear, that the change 
of the effective number of degrees of freedom can occur virtually only and, if 
$\lambda _{n>0}$ state is accompanied  by 
some kind of  equilibration process. 
Both of these statements become clear,
 if we recall that  
the statistical operator in statistical mechanics and the  quantum mechanical convolution operator 
are related by the Wick rotation \cite{Feynmann}. In other words, the inverse temperature can be
considered as an  imaginary time.  
Therefore, depending on the sign,  the quantity  $ I_n b T \equiv \tau_n^{-1}$  that  appears 
 in the trigonometric  functions      
of  the  equations (\ref{eleven}) and (\ref{twelve}) in front of the imaginary time $1/T$ 
can be regarded  as the inverse decay/formation time $\tau_n$ of the metastable state which corresponds to the  pole $\lambda _{n>0}$ (for more details see next sections).
}
 
This interpretation of  $\tau_n$ naturally explains the thermodynamic  metastability 
of all states except the gaseous one:
the metastable states can exist  in the system only virtually 
because of their finite decay/formation  time,
whereas the gaseous state is stable because it has an infinite decay/formation time.

%
%



\vspace*{-0.2cm}

\section{No Phase Transition Case}

\vspace*{-0.2cm}

It is instructive to treat the effective chemical potential $\nu (\lambda)$ as an independent variable
instead of $\mu$. In contrast to the infinite $V$, where  the upper   limit  $\nu \le 0$ defines the liquid phase singularity of the
isobaric partition and  gives the pressure of a liquid phase
$p_l(T,\mu) = T R_0 |_{V \rightarrow \infty}  = (\mu + W(T))/b$ \cite{Bugaev:00,Bugaev:01}, 
for finite  volumes and finite $K(V)$ the effective  chemical potential can
be complex (with either sign for its real part)  and its value defines the number and position of the imaginary roots 
$\{\lambda _{n>0} \}$ 
in the complex plane.
Positive  and negative values of the effective chemical potential  for finite systems  were  considered 
\cite{Elliott:01}
within the Fisher droplet model, but, to our knowledge,  its complex values have never  been  discussed.
From the definition of the effective chemical potential $\nu(\lambda)$ it is evident that  its complex
values for finite systems  exist  only  because of the excluded volume interaction, which is 
not taken into account in the  Fisher droplet model \cite{Fisher:67}.
However, a recent study  of  clusters of 
the $d = 2$ Ising model within the framework of  FDM (see the corresponding section 
in Ref.  \cite{Elliott:05wci})
shows that the excluded volume correction improves 
essentially the description of the thermodynamic functions. 
Therefore, the next step is to consider the complex values of the
effective chemical potential and free energy for the excluded volume correction of the Ising
clusters on finite lattices. 

As it is seen from the upper panel of  Fig.~4, the r.h.s. of Eq. (\ref{twelve})  
is the amplitude and frequency modulated sine-like  
function of dimensionless parameter $I_n\,b$. 
Therefore, depending on $T$ and $Re(\nu)$ values, there may exist 
no 
complex roots $\{\lambda _{n>0}\}$, a finite number of them, or an infinite number of them. 
In the upper panel of  Fig.~4 we showed a special case which corresponds to  exactly three 
roots of Eq. (\ref{ten}) for each value of $K(V)$: the real root ($I_0 = 0$) and two complex conjugate
roots ($\pm I_1$). 
Since 
the r.h.s. of (\ref{twelve}) is monotonously increasing function
of  $Re(\nu)$, when the former is positive,  
it is possible to map the $T-Re(\nu)$ plane into
regions of a fixed number of roots of Eq. (\ref{ten}). 
Each curve in the lower panel of \mbox{Fig.~4} divides the $T-Re(\nu)$ plane
into three parts: for $Re(\nu)$-values below the curve there  is only one real root (gaseous phase), 
for points on  the curve there exist      
three roots, and above the curve there are four or more roots of Eq. (\ref{ten}).

For constant values of  $K(V) \equiv K$  the number of terms in the r.h.s. of (\ref{twelve}) does not depend on
the volume and, consequently, in thermodynamic limit $V \rightarrow \infty$   only the 
rightmost  simple pole in the complex $\lambda$-plane survives out of a finite number of simple poles.
According to the  inequality (\ref{thirteen}), the real root $\lambda_0$ is  the rightmost singularity of isobaric partition (\ref{six}).
However,  there is a  possibility  that the real parts  of other  roots $\lambda_{n>0} $ become infinitesimally 
close to $R_0$, when there is an infinite number of terms which contribute to the GCP (\ref{nine}).

Let us show  now that even for an infinite number of simple poles in (\ref{nine})
only the real root $\lambda_0$  survives in the limit $V \rightarrow \infty$.
For this purpose consider  the limit
$Re(\nu_n)  \gg T $.
In this limit   the distance between  the imaginary parts of the nearest roots 
remains finite even for infinite volume.  Indeed,  for $Re(\nu_n)  \gg T   $
the leading contribution to the r.h.s. of (\ref{twelve}) corresponds to the harmonic with $k = K$,
and, consequently,  an exponentially large amplitude of this term
can be only  compensated by  a vanishing value of  $\sin\left( I_n \, b K  \right)$,  
i.e.   $I_n \, b K  =  \pi n + \delta_n$
{  
with $|\delta_n| \ll \pi$ (hereafter we will analyze only 
the branch $I_n > 0$), 
and, therefore, the corresponding decay/formation time $\tau_n  \approx K [ \pi n T ]^{-1}$ 
is volume independent.

Keeping the leading term on the r.h.s. of  (\ref{twelve}) and solving for $\delta_n$, one finds
\begin{eqnarray}\label{Mfourteen}
\hspace*{-0.2cm}
I_n & \approx & (-1)^{n+1}  \tilde\phi_K (T)  ~{\textstyle e^{\frac{Re(\nu_n )\,K}{T} } } ~\delta_n \,, 
\quad {\rm with} \quad 
%
\delta_n \approx    \frac{ (-1)^{n+1}  \pi n }{ K b ~ \tilde\phi_K (T)  }~{
\textstyle e^{- \frac{Re(\nu_n )\,K}{T} } } \,,    \\
\label{Msixteen}
R_n & \approx & (-1)^{n}  \tilde\phi_K (T)  ~{\textstyle e^{\frac{Re(\nu_n )\,K}{T} } }  \,, 
\end{eqnarray}
where in the last step we used Eq. (\ref{eleven}) and condition $|\delta_n| \ll \pi$.
Since for $V \rightarrow \infty$ all  negative values of $R_n$ cannot contribute to the 
GCP (\ref{nine}), it is sufficient to analyze even values of $n$ which, according to 
(\ref{Msixteen}),  generate  $R_n > 0$.

Since the inequality  (\ref{thirteen}) can not be broken,   a single possibility,
when $\lambda_{n>0}$ pole can contribute to the partition (\ref{nine}),  corresponds to 
the case 
$ R_n \rightarrow R_0 - 0^+$ 
for  some finite $n$.   
Assuming this, we find  $Re (\nu (\lambda_n)) \rightarrow Re(\nu (\lambda_0))$
for the same value of $\mu$. 
}
Substituting these results into equation (\ref{eleven}), one gets
\begin{equation}\label{Mseventeen}
\hspace*{-0.2cm}R_n \approx  \sum\limits_{k=1}^{K } \tilde\phi_k (T)
~{\textstyle e^{\frac{Re(\nu (\lambda_0) )\,k}{T} } } \cos\left[ \frac{ \pi n k}{K} \right]  \ll R_0\,.
\end{equation}
The inequality (\ref{Mseventeen})  follows  from the equation for $R_0$ and the fact that, even for
equal leading terms   in the sums above (with $k =  K$ and even  $n$),  the difference between $R_0$  and $R_n$ is  large due to  the next to leading term $k = K - 1$, which is  proportional to 
$e^{\frac{Re(\nu (\lambda_0) )\,(K-1)}{T} } \gg 1$.  
Thus, we arrive at  a  contradiction with our assumption $R_0 - R_n \rightarrow 0^+$, 
and, consequently,  it cannot be true.  Therefore,
for large volumes  the  real root $\lambda_0$  always gives
the main contribution to  the GCP (\ref{nine}), and  this is the only root that survives 
in the limit $V \rightarrow \infty$.
Thus,  we showed that  the model with the fixed  size  of the largest fragment has no phase transition because there is a single singularity of the isobaric partition (\ref{six}), which 
exists in thermodynamic limit.
 
\vspace*{-0.3cm}
 
\section{Finite Volume Analogs of Phases}

\vspace*{-0.2cm}

If $K(V)$ monotonically grows with the volume,  the situation is different. 
In this case for  positive value of $Re(\nu)  \gg T$ 
the leading exponent in the r.h.s. of (\ref{twelve})  
also corresponds to a largest fragment, i.e. to $k  = K(V)$. 
Therefore,  we can apply 
the same arguments  which were used above  for  the case $K(V) = K =  const$
and derive similarly  equations  (\ref{Mfourteen})--(\ref{Msixteen}) for  $I_n$ and $R_n$. 
From $I_n  \approx \frac{\pi n}{ b\, K( V) }$ it follows that,
when $V$ increases, the number of simple poles in (\ref{eight}) also increases  and
the  imaginary part of 
the closest to the real $\lambda$-axis  poles becomes very small,
 i.e $I_n   \rightarrow 0$ for  $n \ll K(V)$,
 and, consequently, the associated   decay/formation time 
$\tau_n  \approx K(V)  [ \pi n T ]^{-1}$ grows with the volume of the system.
Due to $ I_n  \rightarrow 0$, 
the inequality (\ref{Mseventeen}) cannot be  
established for the poles with $n \ll K(V)$. 
Therefore, in  contrast to the previous case, for large $K(V)$ the  simple poles
with $n \ll K(V)$ will be infinitesimally close to the real axis of the complex $\lambda$-plane.

From Eq.  (\ref{Msixteen}) it follows that 
\vspace*{-0.2cm}
\begin{equation}\label{Meighteen}
R_n ~ \approx ~  \frac{p_l(T,\mu) }{T} -  \frac{ 1}{ K(V) b} 
 \ln \left|  \frac{ R_n}{  \tilde\phi_K (T)     }  \right|  \rightarrow  \frac{p_l(T,\mu) }{T} 
\end{equation}
for  $ | \mu | \gg T $ and $K(V) \rightarrow \infty$.   
Thus, 
we proved that
for infinite volume the  infinite number of simple poles moves toward 
the real $\lambda$-axis to the vicinity of liquid phase singularity $\lambda_l = p_l(T,\mu)/T $ 
of the isobaric partition
\cite{Bugaev:00, Bugaev:01} and
generates  an essential singularity of function ${\cal F}(V, p_l/T)$ in (\ref{seven}) 
{\it irrespective to the  sign of 
the chemical potential $\mu$.}
In addition,  as  we showed above, the states with $Re( \nu ) \gg T$  become  stable because they acquire   infinitely large 
decay/formation time $\tau_n$ in the limit $V \rightarrow \infty$.  Therefore, these states should be identified 
as a liquid phase for finite  volumes as well.

Now it is clear 
that each curve in the lower panel of Fig.~4  is   the  finite volume analog of the phase boundary $T-\mu$ for a given value of $K(V)$:  below the phase boundary there exists a gaseous phase, but at and above each curve there
are  states which can be identified with a finite volume analog of the mixed phase, and,
finally, at $ Re(\nu) \gg T$ there exists a liquid phase.
When  there is no phase transition, i.e. $K(V) = K = const$,  the structure of simple poles is
similar, but, first,  the line which separates the gaseous states from the metastable states does not
change with the volume, and, second, as shown above, the metastable states will never become
stable. 
Therefore,
a systematic study of the 
volume dependence  of free energy (or pressure for very large  $V$)  along with the formation and  decay times may  be  of a crucial importance for  experimental studies of 
the nuclear liquid gas phase transition.

The above results demonstrate that, in contrast  to Hill's expectations \cite{Hill}, the finite volume analog
of the mixed phase does not  
{ consist  just of  two pure phases. }
The mixed phase for finite volumes
consists of  a stable  gaseous phase and  the set of  metastable states which differ by the free energy. 
Moreover, the difference between the free energies of these states is  not surface-like, as Hill 
assumed in his treatment \cite{Hill}, but  volume-like.  Furthermore, 
according to Eqs. (\ref{eleven}) and (\ref{twelve}),  each of these states 
consists of the same fragments, but with different weights. 
As  seen above for the case $ Re(\nu) \gg T$,  
some fragments 
that  belong to 
the states, in which  the largest fragment is  dominant,
may, in principle, have negative weights (effective number of degrees of freedom) in 
the expression  for  $R_{n>0}$  (\ref{eleven}).
This can be understood easily because  higher concentrations of large fragments can be achieved 
 at the  expense of the  smaller fragments and  is reflected in  the corresponding change 
of the real part of  the free energy $- R_{n>0} V T$. 
Therefore, the  actual  structure of the mixed phase at finite volumes is  more complicated
than  was expected in earlier works.  

{
The Hills'  ideas were developed further in Ref.  \cite{Chomaz:03}, where the authors claimed to 
establish the one to one correspondence between the bimodal structure of the partition of measurable
quantity  $B  $ known on average and  the properties of the Lee-Yang zeros of this 
partition in the complex  $  g $-plane. The starting point of  Ref.  \cite{Chomaz:03} is to 
postulate the partition $Z_{ g} $ and  the probability $P_{ g} (  B  )$ of  the following form
\begin{equation}\label{Mnineteen}
Z_{ g}~  \equiv~ \int d { B}~ W ( { B} ) ~ e^{ - { B} \cdot { g}  }  \quad
 \Rightarrow  \quad 
 P_{ g} (  B )~ \equiv ~ \frac{  W ( { B} ) ~ e^{ - { B} \cdot { g}   }  }{  Z_{ g} }\,,
\end{equation}
where $ W ( { B} )$ is the partition  sum of the ensemble of fixed values of the observable  $\{  B \}$ , and
$  g $ is the corresponding Lagrange multiplier. Then the authors of  Ref.  \cite{Chomaz:03}
assume 
the existence of two maxima of the probability  $P_{ g} ( B )$ ($\equiv$ bimodality)  and 
discuss their relation to the Lee-Yang zeros of  $Z_{ g} $  in the complex $ g $-plane.

The CSMM  gives us a unique opportunity  to verify the Chomaz and Gulminelli idea on the bimodality behavior of  $P_{ g} (  B ) $ using the first principle results. Let us use the equation (\ref{nfive})  identifying  the intensive variable $ g $ with  $  \lambda $ and extensive one  $ B  $ with
the available volume $V^\prime \rightarrow \tilde V$. The evaluation of the r.h.s. of  (\ref{nfive}) 
is very difficult in general, but for a special case, when the eigen volume $b$ is small this can be done
analytically. Thus, approximating ${\cal F}(\xi, \lambda - i \eta) \approx {\cal F}(\xi, \lambda) -
i\eta \, \partial {\cal F}(\xi, \lambda)/ \partial \lambda $, we obtain the CSMM analog of  the probability 
(\ref{Mnineteen})
\begin{equation}\label{Mtwenty}
 P_{\lambda} ( \tilde V ) ~ \hat{\cal Z}(\lambda,T,\mu) \equiv  \int\limits_{-\infty}^{+\infty} d \xi~ 
 \int\limits_{-\infty}^{+\infty}
  \frac{d \eta}{{2 \pi}} ~ { \textstyle e^{ i \eta ( \tilde V - \xi) - \lambda \tilde V
+ \tilde V {\cal F}(\xi, \lambda - i \eta) } } \approx  
\int\limits_{-\infty}^{+\infty} d \xi~  { \textstyle e^{ \tilde V  [   {\cal F}(\xi, \lambda )   -\lambda ]  } 
\delta\left[ \tilde V - \xi -    \frac{\partial {\cal F}(\xi, \lambda ) }{\partial \lambda}   \right]  }
\,,
\end{equation}

\newpage
\noindent
where we made the $\eta$ integration after  applying the approximation for 
 ${\cal F}(\xi, \lambda - i \eta)$. Further evaluation of (\ref{Mtwenty})  requires to know all 
 possible solutions of the average  volume of the system 
 $\xi^*_\alpha ( \tilde V ) = \tilde V -  \partial {\cal F}(\xi^*_\alpha, \lambda)/ \partial \lambda $
  ($\alpha = \{1,2,\dots\}$). It can be shown 
 \cite{Bugaev:05csmm}
 that for the gaseous domain $\nu = Re( \nu) < - 2 T$ (see the lower panel of Fig. 4)  there exist a single solution
 $\alpha = 1$,
 whereas for the domain $ \nu = Re( \nu) > 0 $, which corresponds to a  finite volume analog of  the mixed phase,   there are
 two solutions $\alpha= 1, 2$.  In contrast to the expectations of  Ref. \cite{Chomaz:03},
 the probability   (\ref{Mtwenty})  
\begin{equation}\label{Mtwone}
 P_{\lambda} ( \tilde V ) ~ \hat{\cal Z}(\lambda,T,\mu) \approx 
 \sum\limits_{ \alpha }  \frac{ 1 }{ \left|  1 + \frac{\partial^2  {\cal F}(\xi^*_\alpha, \lambda)  }{\partial \lambda \xi^*_\alpha}  \right|  }    { \textstyle e^{ \tilde V  [   {\cal F}(\xi, \lambda )   -\lambda ]  } }  \quad 
 \Rightarrow  \quad  \frac{ \partial \ln P_{\lambda} ( \tilde V )  }{ \partial   \tilde V } ~\le~ 0
 \,,
\end{equation}
 has  negative  derivative for the whole domain  of existence of the  isobaric partition 
 $\hat{\cal Z}(\lambda,T,\mu)$  \cite{Bugaev:05csmm}.
 This is true  even for   the domain in which, as we proved,  there exists  a finite analog
 of the mixed phase, i.e. for  $ \nu = Re( \nu) > 0 $. 
 Moreover, irrespective  to  the sign of  the derivative 
 $ \frac{ \partial \ln P_{\lambda} ( \tilde V )  }{ \partial   \tilde V }$,   the probability 
 (\ref{Mtwenty}) cannot be measured experimentally.  
 Above it  was rigorously proven  that for
 any  real $\xi$ the IP 
 $\hat{\cal Z}(\lambda,T,\mu)$  is  defined on the real $\lambda$-axis only  for 
 $  {\cal F}(\xi, \lambda )   -\lambda > 0 $, i.e.  on the right hand side  of the gaseous singularity  
 $ \lambda_0$:  $ \lambda > \lambda_0$. However, as one can see from the  equation (\ref{eight}), the ``experimental'' $\lambda_n$
 values belong to the other region of the complex $\lambda$-plane: $ Re( \lambda_{n > 0}) < \lambda_0$.

Thus, it turns out that  the suggestion of   Ref.  \cite{Chomaz:03} to analyze the  probability 
(\ref{Mnineteen}) does not make any sense because, as we showed explicitly for the CSMM, it cannot be measured. 
It seems that   the starting point of  the Ref.  \cite{Chomaz:03}  approach,
i.e.  the  assumption that the left  equation  (\ref{Mnineteen}) gives the most general form of
the partition of  finite system, is problematic. 
Indeed, comparing   (\ref{Meighteen})  with  the analytical result
(\ref{Mtwone}), we see that for finite systems, in contrast to the major  assumption of 
Ref.  \cite{Chomaz:03}, 
the probability  $ W $ of  the  CSMM depends not only on the extensive variable $\tilde V$, but also 
on the intensive variable $\lambda$, which makes unmeasurable  the whole construct of Ref.  \cite{Chomaz:03}. 
Consequently, the conclusions of 
Ref.  \cite{Chomaz:03} on the relation between the bimodality and the phase transition existence  are not general and have a limited range of validity. 
In addition,  the absence of two maxima of the probability  (\ref{Mtwone}) automatically means the
absence of  back-banding  of the equation of state  \cite{Chomaz:03}.  

}



\vspace*{-0.3cm}

\section{Gas of Bags in Finite Volumes}

\vspace*{-0.2cm}

Now we will apply the formalism of the preceding sections to the 
analysis of the Gas of Bags Model (GBM) \cite{Goren:81,Goren:05}  in finite volumes. 
In the high and low temperature domains 
the GBM  reduces to two well known and
successful  models:
the hadron gas model  \cite{Hgas}  and the bag model of QGP \cite{BagModel}.
Both of these models are surprisingly successful in describing the bulk properties 
of hadron production in high energy nuclear collisions, and,
therefore, one may hope  that  their generalization, the GBM, may reflect basic features of the nature  
in the  phase transition region.

The van der Waals gas consisting of $n$ hadronic  species,
which are called bags in what follows,  has the following  GCP \cite{Goren:81}
\begin{equation}
 Z  (V,T)  =  
 \sum_{\{N_k\}} \biggl[
\prod_{k=1}^{n}\frac{\left[ \left( V -v_1N_1-...-v_nN_n\right)  ~\phi_k(T) \right]^{N_k}}{N_k!} \biggr] 
%
~ \theta\left(V-v_1N_1-...-v_nN_n\right)~, 
\label{qqq}
\end{equation}
\vspace*{-0.2cm}

\noindent
where $\phi_k(T) \equiv g_k ~ \phi(T,m_k)  \equiv  \frac{g_k}{2\pi^2}~\int_0^{\infty}p^2dp~
\exp\left[-~  (p^2~+~m_k^2)^{1/2} / T \right]
~=~  g_k \frac{m_k^2T}{2\pi^2}~K_2\left( \frac{m_k}{T} \right)$  is the particle  density
of  bags of mass $m_k$ and eigen volume $v_k$  and degeneracy $g_k$. 
This expression differs  slightly  form the GCP of the simplified SMM  (\ref{three}), where 
$\mu = 0$ and  the  eigen volume of $k$-nucleon fragment $k b$ is changed to 
the  eigen volume of the bag $v_k$.
Therefore,  as  for  the simplified SMM the Laplace transformation  (\ref{four})   with respect to volume  of Eq.~(\ref{qqq}) gives 
\vspace*{-0.2cm}

\begin{equation}\label{Zsn}
 \hat{Z}  (s,T)~=~\left[~s~-~\sum_{j=1}^n \exp\left(-v_j s\right)~g_j\phi(T,m_j)\right]^{-1}~.
 \end{equation}
 \vspace*{-0.3cm}

\noindent
In preceding sections we showed that 
as long as the number of bags, $n$, is finite, the only possible singularities  
of $\hat{Z}(s,T)$ (\ref{Zsn}) are simple   poles. 
However, in the case of an infinite number of bags  an essential  singularity of
$\hat{Z}(s,T)$ may appear.  
This  property is used  the GBM: 
the sum  over different bag states in (\ref{qqq})
can be  replaced  by the integral,
$\sum_{j=1}^{\infty}g_j ...=\int_0^{\infty}dm\, dv ...\rho(m,v)$, 
if   the bag mass-volume spectrum, $\rho(m,v)$,  which defines  
the number of bag states in the mass-volume  region  $[m,v;m+dm,v+dv]$,
is given.  Then, the Laplace transform of $Z(V,T)$ reads \cite{Goren:81}
\vspace*{-0.5cm}

\begin{equation}\label{Zsbag}
 \hat{Z}_{GB} (s,T) \equiv \int\limits_0^{\infty}dV ~ e^{-sV}~Z(V,T)
= \left[~s~-~f(T,s)\right]^{-1}\,,~
{\rm where} \quad
 f(T,s)=  \int\limits_0^{\infty} dm\,dv ~\rho(m,v)~e^{-vs}~\phi(T,m)\,.
\end{equation}
\vspace*{-0.3cm}

\noindent
Like in the simplified SMM, 
the  pressure  of infinite system is again given by the rightmost singularity: 
$p(T)=Ts^*(T)~=~T\cdot max\{s_H(T),s_Q(T)\}$.  
Similarly to the simplified SMM considered in Sect. II and III, the rightmost  singularity $s^*(T)$ of $\hat{Z}(s,T)$ (\ref{Zsbag})
can be either the simple  pole singularity $s_H(T) ~=~f\left(T,s_H(T)\right) $ of the isobaric partition
(\ref{Zsbag})  or 
the $s_Q(T)$ singularity of the function $f(T,s)$ (\ref{Zsbag}) it-self.

The major mathematical difference between the simplified SMM and the GBM is that the latter 
employs the two parameters  mass-volume spectrum.  Thus, the mass-volume spectrum of the GBM
consists of the  discrete mass-volume spectrum of light hadrons and the continuum contribution
of heavy resonances \cite{Goren:82}
\vspace*{-0.2cm}
\begin{equation}\label{rhomv}
 \rho(m,v) =  \sum_{j=1}^{J_m}~ g_j~ \delta(m-m_j)~\delta(v-v_j) +  \Theta(v -V_0) \Theta(m -M_0 -Bv) C~v^{\gamma}(m-Bv)^{\delta} ~\exp\left[\frac{4}{3}~\sigma_Q^{  \frac{1}{4} }~
 v^{ \frac{1}{4} }~(m-Bv)^{\frac{3}{4} }\right]~,
\end{equation}
\vspace*{-0.3cm}

\noindent
respectively.  Here $m_j < M_0$, $ v_j < V_0$,  $M_0 \approx 2 $ GeV, $V_0 \approx 1$ fm$^3$,
$C, \gamma, \delta$ and $B$ (the so-called bag constant, $B \approx 400$ MeV/fm$^3$) are 
the model parameters and 
\begin{equation}\label{sigmaQ}
 \sigma_Q~=~\frac{\pi^2}{30}\left(g_g~+~\frac{7}{8}g_{q\bar{q}}\right)~
 =~
\frac{\pi^2}{30}\left(2\cdot 8~+~\frac{7}{8}\cdot 2\cdot 2\cdot 3
\cdot 3\right)~=~\frac{\pi^2}{30}~\frac{95}{2}
 \end{equation}
\noindent
is the Stefan-Boltzmann constant counting gluons (spin, color) and (anti-)quarks
(spin, color and $u$, $d$, $s$-flavor) degrees of freedom.

Recently the grand canonical ensemble has been heavily criticized 
\cite{HThermostat:1, HThermostat:2}, when it is used for 
the exponential mass spectrum.  This critique, however, cannot be applied to the mass-volume 
spectrum (\ref{rhomv}) because it  grows less fast than the Hagedorn mass spectrum discussed 
in \cite{HThermostat:1, HThermostat:2}  and  because in the GBM  there is 
an additional suppression of  large and heavy bags due to the van der  Waals  repulsion. 
Therefore, the spectrum (\ref{rhomv})  can be  safely used in  the grand canonical ensemble.

It can be shown \cite{Goren:05}  that  the spectrum (\ref{rhomv})  generates  the
$s_Q (T) = \frac{\sigma_Q}{3} T^3 - \frac{B}{T}$ singularity, which reproduces the bag model  pressure 
$p (T) = T s_Q (T)$ \cite{BagModel} for high temperature phase, and $s_H (T)$ singularity,  which  gives   the pressure of the hadron gas model \cite{Hgas} for low temperature phase. The transition between 
them can be of the first order or second order or cross-over, depending on the model parameters.

However, for finite systems the volume of  bags  and their  masses should be finite. 
The simplest finite volume  modification of the GBM  is to introduce the volume dependent 
size of the largest bag $n = n (V)$  in  the partition  (\ref{qqq}).  As we discussed earlier such a modification cannot be handled by the traditional Laplace transform technique used in 
\cite{Goren:82,Goren:05}, but this modification can be easily
accounted for by the Laplace-Fourier method \cite{Bugaev:04a}.  Repeating all the steps 
of the sections V and VI, we shall  obtain the  equations  (\ref{seven})-(\ref{ten}), in which the function 
${\cal F}(\xi,\tilde\lambda)$ should be replaced  by its GBM analog 
$f(\lambda,V_B) \equiv   f_H( \lambda) +  f_Q(\lambda, V_B)$ defined via 
\vspace*{-0.3cm}
\begin{equation}\label{Fgbm}
 f_H( \lambda) \equiv \sum_{j=1}^{J_m}~ g_j \, \phi(T,m_j) ~ e^{-v_j s }\,, \quad  {\rm and} \quad
f_Q(\lambda, V_B) \equiv  V_0  \int\limits_{1}^{V_B/V_0}  \,dk  ~ a (T,V_0 k )~ e^{ V_0 (s_Q(T) - \lambda) k} \,.
\end{equation}
\noindent
In evaluating (\ref{Fgbm}) we used the mass-volume  spectrum (\ref{rhomv}) with  the maximal volume of the bag $V_B$ and changed integration to a dimensionless variable  $k = v / V_0$. Here the function 
$ a (T,v) = u (T) v^{2 + \gamma + \delta}$ is defined by  
$u(T) = C \pi^{-1} \sigma_Q^{\delta + 1/2}~ T^{4 + 4 \delta} ( \sigma_Q T^4 + B )^{3/4}$. 

The above representation  (\ref{Fgbm}) generates  equations for  the real and  imaginary parts of
$\lambda_n \equiv R_n + i I_n$, which are very similar to the corresponding expressions of the CSMM 
(\ref{eleven})  and (\ref{twelve}).  Comparing (\ref{Fgbm}) with (\ref{ten}), 
one sees that their main difference is that the sum over $k$ in (\ref{ten}) is replaced by the integral
over  $k$ in (\ref{Fgbm}).  Therefore, the equations (\ref{eleven})  and (\ref{twelve}) remain valid for
$R_n$ and $I_n$ of the GBM, respectively, if we replace the $k$ sum by the integral for  
$K(V) = V_B/V_0$, $b = V_0$, $\nu (\lambda) =  V_0 ( s_Q (T) - \lambda)$ and 
$\tilde\phi_{k>1} (T) =  V_0  ~ a (T,V_0 k)  $. Thus, the results and  conclusions of
our analysis of the $R_n$ and $I_n$ properties of the CSMM should be  valid for the GBM as well.
In particular, for  large  values of $K(V) = V_B/ V_0$  and $R_n < s_Q(T) $ one can immediately 
find out  $I_n \approx \pi n  / V_B $ and the GBM formation/decay time  
$\tau_n = V_B  [ \pi n T V_0]^{-1}$. These equations show that the metastable $\lambda_{n>0}$ 
states can become stable in thermodynamic limit, if and only if  $V_B \sim V$.

The finite volume modification of the GBM equation of state should be used for the quantities which
have $V \lambda_0  \sim 1$.  This may  be important for  the early stage of   the relativistic nuclear collisons when the volume of the system  is small, or for the  systems that have small pressures. 
The latter can be the case for  the  pressure of strange or charm hadrons.

\vspace*{-0.3cm}

\section{ Hills and Dales Model and the Source of Surface Entropy}

\vspace*{-0.2cm}

During last forty years the Fisher droplet model (FDM) \cite{Fisher:67}
has been successfully  used to analyze the condensation  of 
a gaseous phase (droplets or clusters  of all sizes)    
into a liquid.  
The systems analyzed with the FDM are many and varied, but
up to now the source of the surface entropy is not absolutely clear.
In his original work  Fisher postulated that 
the surface free-energy $F_A$ of a  cluster  of $A$-constituents  consists 
of surface  ($A^{2/3}$)  and  logarithmic ($\ln A$) parts, i.e. 
$F_A =  \sigma (T)~ A^{2/3} + \tau  T\ln A$.  Its surface part  
$ \sigma (T)~ A^{2/3} \equiv \sigma_{\rm o} [ 1~ - ~T/T_c] ~ A^{2/3}$  consists 
of the  surface energy, i.e. $ \sigma_{\rm o}  ~ A^{2/3}$,  and  
surface entropy $ -  \sigma_{\rm o} / T_c~ A^{2/3}$. 
From the study of the combinatorics of lattice gas clusters in two dimensions,
Fisher  postulated    the  specific 
temperature dependence of the surface tension $\sigma (T)|_{\rm FDM} $
which gives 
naturally an estimate   for the  critical temperature  $T_c$. Surprisingly  
Fisher's estimate works  for  the 3-d Ising model  \cite{Ising:clust},  
nucleation of real fluids \cite{Dillmann,Kiang}, percolation  clusters \cite{Percolation}
and  nuclear multifragmentation \cite{Moretto:97}. 

To understand why the surface entropy has such a form we formulated 
a statistical model of surface deformations of the cluster of $A$-constituents, the Hills and Dales Model (HDM)   \cite{Bugaev:04b}. 
For simplicity we consider
 cylindrical deformations of positive height $h_k>0$ (hills) 
and negative height $-h_k$ (dales), with  $k$-constituents at the base. 
It  is assumed that   cylindrical deformations of positive height $h_k>0$ (hills) 
and negative height $-h_k$ (dales), with  $k$-constituents at the base,  and
the top (bottom) of the hill (dale) has the same shape as 
the surface of the original  cluster of $A$-constituents. 
We also  assume that:
(i) the statistical weight of deformations $\exp\left( - \sigma_{\rm o} |\Delta S_k|/s_1 /T  \right) $ 
is given  by the Boltzmann factor due to the  change of the surface $|\Delta S_k|$ in units of 
the surface per  constituent $s_1$;
(ii) all hills of heights $h_k \le H_k$ ($H_k$ is the maximal height of 
a hill with a base of $k$-constituents)
have the same probability $d h_k/ H_k$ besides the statistical one; 
(iii) assumptions (i) and (ii) are valid for the dales.

The HDM grand canonical surface partition (GCSP)
\vspace*{-0.2cm}
\begin{equation} \label{Oone}
Z(S_A)= \hspace*{-0.10cm} \sum\limits_{\{n_k^\pm = 0 \}}^\infty \hspace*{-0.10cm} \left[ \prod_{k=1}^{ K_{max} }
\frac{ \left[ z_k^+ {\cal G} \right]}{n^+_k!}^{n^+_k} \frac{ \left[ z_k^- {\cal G} \right]}{n^-_k!}^{n^-_k}\right]
\Theta(s_1 {\cal G})\, 
\end{equation}
\vspace*{-0.25cm}

\noindent 
corresponds to the conserved (on average) volume of the cluster because the probabilities 
of hill $z_k^+$  and dale   $z_k^-$
of the same $k$-constituent  base  are identical   \cite{Bugaev:04b}
\vspace*{-0.2cm}
\begin{equation}\label{Otwo}
\hspace*{-0.35cm}
z_k^{\pm} \equiv \hspace*{-0.15cm} \int\limits_0^{\pm H_k} \hspace*{-0.15cm} \frac{ d h_k}{ \pm H_k}\,
{\textstyle e^{ - \frac{\sigma_{\rm o} P_k |h_k| }{T s_1} } }
= \frac{T s_1  }{\sigma_{\rm o} P_k H_k }
\left[1 - {\textstyle e^{ - \frac{\sigma_{\rm o} P_k H_k}{ T s_1} } } \right] .
\end{equation}
\vspace*{-0.25cm}

\noindent 
Here $P_k$ is the perimeter of the cylinder base. 

The geometrical partition (degeneracy factor) 
of the HDM 
or the number  of ways to place
the center of a  given  deformation on the surface of the $A$-constituent cluster which  is occupied
by the set of $\{n_l^\pm  = 0, 1, 2,...\}$  deformations of the $l$-constituent base we assume to
be  given in the van der Waals approximation  \cite{Bugaev:04b}:
\vspace*{-0.2cm}
\begin{equation}\label{Othree}
{\cal G} = 
{\textstyle \left[ S_A - \sum\limits_{k = 1}^{K_{max} } k\, (n_k^+ ~ + ~ n_k^-) \, s_1 \right] s_1^{-1} } \,, 
\end{equation}
\vspace*{-0.25cm}

\noindent 
where $s_1 k$ is the area  occupied by the deformation of $k$-constituent base ($k = 1, 2,...$), 
$ S_A$
is the  full surface of the cluster,
and $K_{max} (S_A) $ is the $A$-dependent size of the maximal allowed base on the cluster.

The $\Theta(s_1 {\cal G})$-function in (\ref{one}) ensures that only configurations
with positive value of the free surface of cluster are taken into account, but makes 
the  analytical evaluation of the  GCSP (\ref{one}) very difficult.  However,  we were able {\it  to solve
this GCSP  exactly}  for any surface dependence of $K_{max} (S_A) $  using identity (\ref{nfour}) of  the Laplace-Fourier transform  technique \cite{Bugaev:04a}:
\vspace*{-0.3cm}

\begin{equation}\label{Ofour}
 Z(S_A)~ = \sum_{\{ \lambda_n\}}
e^{\textstyle  \lambda_n\, S_A }
{\textstyle
\left[1 - \frac{\partial {\cal F}(S_A,\lambda_n)}{\partial \lambda_n} \right]^{-1} } \,.
\end{equation}
\vspace*{-0.25cm}

\noindent 
The  poles  $\lambda_n$  of the isochoric partition  are defined by  

\vspace*{-0.3cm}

\begin{equation}\label{Ofive}
\lambda _n~ = ~{\cal F}(S_A,\lambda _n) \equiv  \sum\limits_{k=1}^{ K_{max}(S_A) } 
\left[  \frac{ z_k^+}{s_1} + \frac{ z_k^-}{s_1} \right]
~e^{ - k\,s_1 \lambda_n }
\,.
\end{equation}
\noindent 
Our analysis shows that Eq. (\ref{five})  has exactly one real root 
$R_0 = \lambda_0, Im( \lambda_0) = 0$,
which is the rightmost singularity, i.e. $R_0 > Re( \lambda_{n>0}) $. As  proved in \cite{Bugaev:04b},
the real root  $R_0 $ dominates completely  for  clusters  with  $A \ge 10$.

Also we  showed that there is  an absolute supremum for the real root $R_0$,
which corresponds to  the limit of infinitesimally small amplitudes 
of deformations, $H_k \rightarrow 0$,  of large clusters:
$\sup (R_0) =  1.06009 \equiv R_0 = 2 \left[ e^{ R_0} - 1 \right]^{-1}.$ It is remarkable that
the last result is, first, model independent because
in the limit of vanishing amplitude of deformations all model specific parameters vanish;  and, 
second,  it is valid  for any self-non-intersecting surfaces. 

For large  spherical clusters the GCSP becomes 
$ Z (S_A) \approx  0.3814~ e^{\textstyle  1.06009\, A^{2/3}  } $, which, combined with the Boltzmann factor of
the surface energy $e^{\textstyle - \sigma_{\rm o} A^{2/3}/T  } $, generates
the { following temperature dependent surface tension} of the large cluster 
$
\sigma (T) = \sigma_{\rm o} \left[ 1 - 1.06009 \frac{T}{\sigma_{\rm o} } \right] .
$
This result  means that the actual critical temperature of the
FDM  should be
$T_c = \sigma_{\rm o}/ 1.06009$, i.e. 6.009 \% 
smaller in $\sigma_{\rm o}$ units 
than Fisher originally supposed.

\vspace*{-0.2cm}

\section{Strategy of Success}

\vspace*{-0.2cm}

Here  we  discussed  exact analytical solutions  of a variety of statistical model which 
are obtained by  
a new powerful mathematical method, the Laplace-Fourier  transform.
Using this method we solved the constrained SMM  and  Gas of Bags Model for finite volumes,
and found the surface partition of large clusters. 
Since in the thermodynamic limit  the CSMM has  the  nuclear liquid-gas PT and  the
 GBM  describes the  
PT  between the  hadron gas and QGP, it 
was interesting and important to  study them for finite volumes. 
As we showed,  for finite volumes their  GCP functions 
can be identically rewritten in terms of the simple poles $\lambda_{n \ge 0}$  of the isobaric partition (\ref{six}).
We proved that the real pole $\lambda_0$  exists always and  
the quantity  $T \lambda_0$ is the constrained
grand canonical  pressure of the gaseous phase.  
The complex roots $\lambda_{n>0}$ appear as pairs of complex conjugate solutions
of equation (\ref{ten}).   Their  most straightforward interpretation  is as follows: 
$- T Re (\lambda_{n}) $ has a meaning of 
the free energy density, whereas $ b T Im (\lambda_{n} )$, depending on its  sign,  gives
the  inverse  decay/formation time  of such a state. Therefore, 
the gaseous state is always stable  because its decay/formation time is infinite and because
it has the smallest value of  free energy, whereas 
the complex poles describe the metastable states for $Re( \lambda_{n>0} ) \ge  0 $ and mechanically
unstable states for $Re( \lambda_{n>0} ) < 0 $.

We studied the volume dependence of the simple poles and found a dramatic difference
in their  behavior  for the  case with  phase transition and  without it.
For the case with  phase transition 
this formalism   allows one to define the finite volume analogs of phases  unambiguously 
and to establish the finite volume analog of the $T-\mu$  phase diagram (see Fig. 4). 
At finite volumes the gaseous phase  is  described by  a  simple pole $\lambda_0$, the mixed phase corresponds to a finite number of simple poles (three and more), 
whereas the liquid is represented by an infinite amount of simple poles at  $|\mu|  \rightarrow \infty$
which describe  the states of a  highest possible 
particle density.

As we showed for the  CSMM and GBM,  at finite volumes 
the  $\lambda_n$ states of the same partition with given $T$ and $\mu$  
are
not in a true chemical equilibrium because the interaction between the constituents generates
complex values of the effective chemical potential. This feature cannot be obtained within 
the Fisher droplet model due to lack
of the hard core repulsion between the  constituents.  We showed   that, 
in contrast to Hill's expectations \cite{Hill}, the mixed phase at finite volumes  is  
not just  a composition of two 
states which are the pure phases. 
As we showed, a finite volume analog of the  mixed phase is a superposition of three and more collective states, 
and each of them  is characterized by its own  value of $\lambda_n$, and,  consequently,
the difference between the  free energies  of these states is not a surface-like,
as Hill argued \cite{Hill}, but volume-like.

Also the exact analytical formulas gave us a unique opportunity to verify  the 
Chomaz and Gulminelli  ideas  \cite{Chomaz:03}  about the connection between 
bimodality and the phase transition existence for finite volumes.  
The CSMM exact analytical solution not only provided us with a counterexample 
for which there is no bimodality in case of finite volume phase transition,  but 
it gave us an explicit example to illustrate that the probability which, according to
Ref.   \cite{Chomaz:03} is supposed to signal the bimodal behavior of the system,
cannot be measured experimentally. 

All this clearly demonstrates that  the exactly solvable models are very useful theoretical 
tools and they open the new possibilities to study the critical phenomena at finite volumes
rigorously. The {\bf short range perspectives (SRP)} of this direction of research  are evident:
\begin{enumerate}
\item {\bf Study} the isobaric ensemble and  the excluded volume correction  for 
the clusters of  the 2- and 3-dimensional Ising models, and find out  the reliable signals of phase transition on finite lattices. 

\item {\bf Widen or refine}  the CSMM and GMB  analytical solutions  for more realistic interaction between the constituents. In particular,  a more realistic Coulomb interaction between
nuclear fragments (not the Wigner-Seitz one!)  can  be readily  included  now into the CSMM and 
may  be studied rigorously without taking  thermodynamic limit. 

\item {\bf Deepen or extend}  the CSMM and GMB  models to the canonical and microcanonical formulations, and work out  the reliable signals  of the finite system  phase transitions 
for  this class of models.

\end{enumerate}

The major goals for the SRP are ({\bf I})  to get the reliable experimental signals obtained not with 
the {\it ad hoc} theoretical constructs which are very popular nowadays, but directly from the first principles of statistical mechanics; 
({\bf II}) to work out a common and useful theoretical language for  a few  nuclear physics communities.

However, even the present (very limited!) amount of exact results  can be used as a good starting point
to build up a truly microscopic theory of phase transitions in finite systems. 
In fact,  the exact analytical solution, which we found  for finite volumes,  is 
one of the key elements that are necessary to create a microscopic kinetics of PTs in finite systems. 
The formulation of such a theory for nuclear physics is demanded by the reality of 
the experimental  measurements: 
both of the  phase transitions which are studied in nuclear  laboratories, the liquid-gas and 
hadron gas - QGP, are accessible only via the violent nuclear collisions.  As a result,  in these
collisions we are dealing with the PTs which occur not only in finite system, but in addition 
these PTs  happen dynamically.  
It is  known that  during the course of collision the system experiences a complicated 
evolution from  a highly excited (on the ordinary level) state which is far from local equilibrium,
to the collective expansion of the locally  thermalized  state and to a (nearly)  free-streaming  stage of corresponding constituents. 

A tremendous complexity of  the nuclear collision process makes it extremely difficult for theoretical
modeling. This is, in part, one of the reasons why, despite a great amount of experimental data collected
during last 25 years and numerous theoretical attempts,
neither  the liquid-gas nor  the  hadron gas - QGP phase transitions  are well established experimentally
and well understood theoretically.   It turns out that the major problem of modeling both of these PTs in dynamic situations   is the  absence of  the suitable  theoretical apparatus. 

For example, it is widely  believed in the Relativistic Heavy Ion community (RHIc) that relativistic hydrodynamics
is the best theoretical tool to model the PT between QGP and hadron gas because it employs only 
the equilibrium  equation of state \cite{Heinz:05}.  Up to now this is just a wishful  illusion because 
besides the incorrect boundary conditions, known as {\it freeze-out  procedure''} 
\cite{Bugaev:96, Bugaev:99}, 
which are typically  used in the actual hydro calculations  \cite{Heinz:05}, the employed  equation of state does not fit into the finite (and sometimes small!) size of the system because it corresponds to an infinite system. On the other  hand it is known \cite{Rischke1}  that  hydrodynamic description is limited by the weak (small) gradients of the hydro variables, which define a characteristic scale 
not only for collective hydro properties, but also a typical volume for the equation of state.  

Above we showed  that for finite systems the equation of state inevitably  includes the volume dependence of such  thermodynamic variables  as pressure and  energy density which 
are directly involved into hydrodynamic equations. This simple  fact is not realized yet in the
RHIc, but, probably, the chemical non-equilibrium (which is usually implemented into
equation of state by hand) is, in part, generated by  the finite volume corrections  of the GCP. 
If this  is the case, then, according to our analysis of the finite volume  GCP functions,  
it is necessary to insert  the complex values of the chemical potential into  hydro calculations. 

Unfortunately, at present there is no  safe recipe on  how  to include the finite volume equation 
of state in the hydrodynamic description. A partial success of the hybrid hydro-cascade  models
\cite{BD:00,TLS:01}, which might be considered as a good alternative to hydrodynamics,  
is compensated by the fact that  none  of the existing hydro-cascade  models 
was able to  resolve the so called {\it  HBT puzzles}  \cite{QM:04} found in the  energy range  of  the Relativistic Heavy Ion Collider.
Moreover, despite the rigorous derivation \cite{Bugaev:02HC,Bugaev:04HC}  of the hydro-cascade  equations,
the hydrodynamic part of this approach is suffering the very same problems of the infinite matter equation of state which we discussed above.  
Therefore,  further refinements  of the hydro-cascade  models  
will not be able to  lift  up  the theoretical apparatus
of modeling  the dynamics of  the finite volume  PTs  to  new heights, and we have to search for
a more elaborate approach.  

It turns out  that  the recently derived  
finite domain kinetic equations \cite{Bugaev:02HC,Bugaev:04HC}  
can provide us with  another  starting point to develop the first principle 
microscopic theory of the critical phenomena in finite systems. 
These equations generalize the relativistic Boltzmann equation to finite domains and,
on one hand, 
allow one to conjugate  two  (different!) kinetics which exist in two domains separated by 
the evolving  boundary,  
and, on the other hand,  to  account exactly  for the exchange
of particles between these domains. 
(For instance,  one can easily imagine the situation when on one side of the  boundary separating the domains
there may exist 
one phase of the system  which  interacts  with the other phase located on the other side of 
the boundary.)
But, first, 
the  finite domain  kinetic  equations  should be generalized to the two-particle distribution functions
and then they should be adapted to the framework of nuclear multifragmentation and 
the Gas of Bags Model. 
In doing this, the exact analytical results we discussed will be indispensable 
because they  provide us with the equilibrium state of  the finite system and tell us to what
finite volume analog of  phases this state  belongs. 

Therefore, a future  success in building up a microscopic kinetics of PTs in finite systems 
can be achieved, if we combine the exact results obtained for equilibrated finite systems
with the rigorous  kinetic equations suited for finite systems. 
There is a  good chance for the nuclear multifragmentation community to play 
a very  special role in the development  of  such a theory, namely it may 
{\it  act  as a perfect and  reliable  test site}  to work out and verify the whole  concept. 
This  is so because
besides some theoretical advances and experience in studying the PTs in finite systems, 
the experiments 
at intermediate energies,  compared to the searches for  QGP,  are easier and cheaper to perform, and  
the PT signals  are cleaner  and  unspoiled by a strong flow.
Moreover, once the concept is developed and verified, 
it can be modified and  applied to study other PTs in finite systems,
including the transitions to/from high temperature QCD and dense  hadronic matter 
planned to be studied  at  CERN LHC   and  GSI FAIR. 
Thus, after some readjustment  the manpower and experimental facilities of nuclear 
multifragmentation community can be used for {\it  a new strategic aim,}   which is  at  the frontier  line of 
modern physics.


\vskip3mm

The fruitful  discussions with  J. B. Elliott,  L.G. Moretto  and L. Phair  are appreciated.
This work was in part supported by the US Department of Energy.
P.T.R. acknowledges the financial support from  the German Academic Exchange Service (DAAD).
%


 \begin{flushright}
{\footnotesize Received 30.10.06}
\end{flushright}


\newpage

\begin{figure}
%

\includegraphics[width=7.0cm,height=5.5cm]{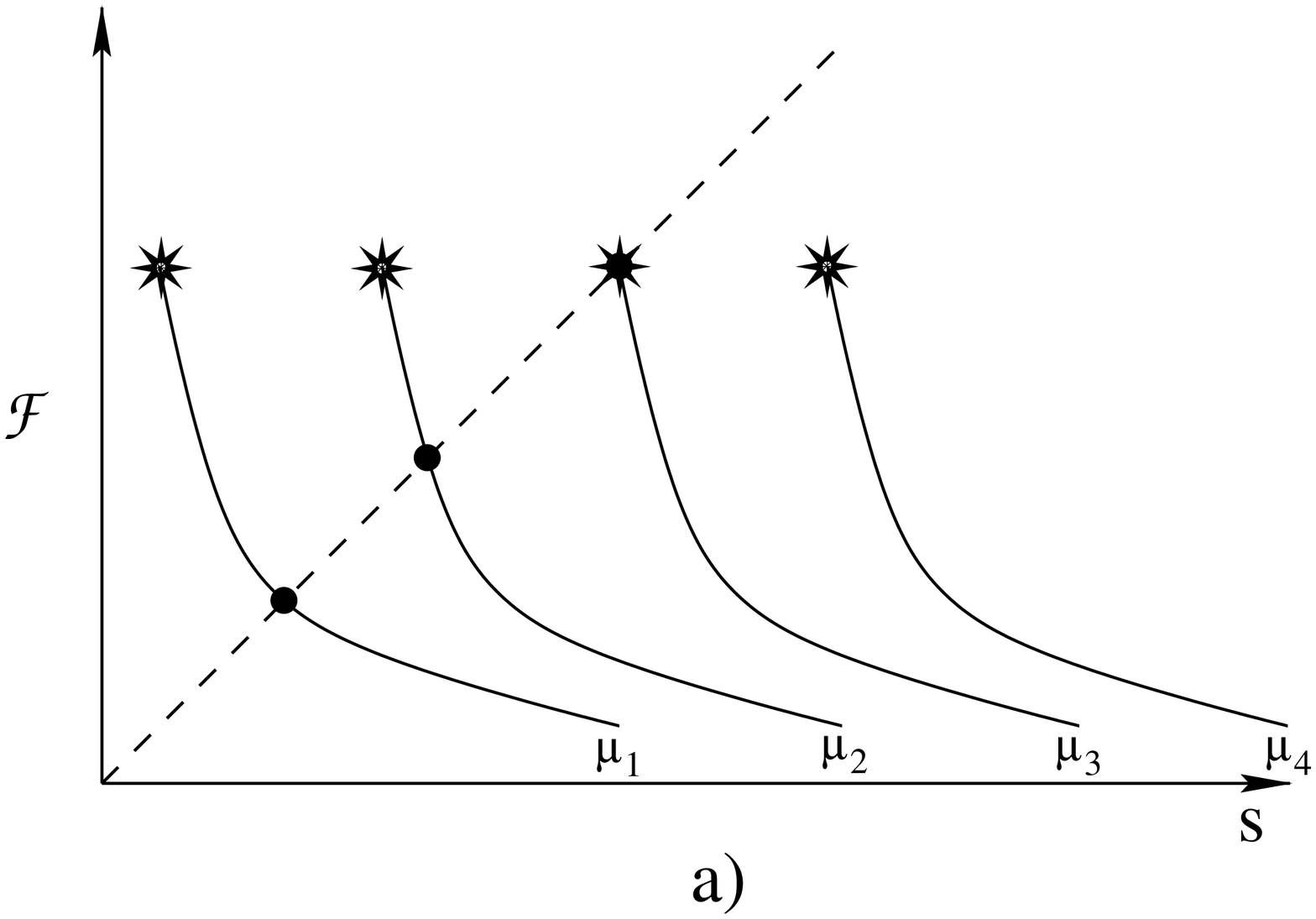}

{
\vspace*{1.0cm}
\includegraphics[width=7.0cm,height=5.0cm]{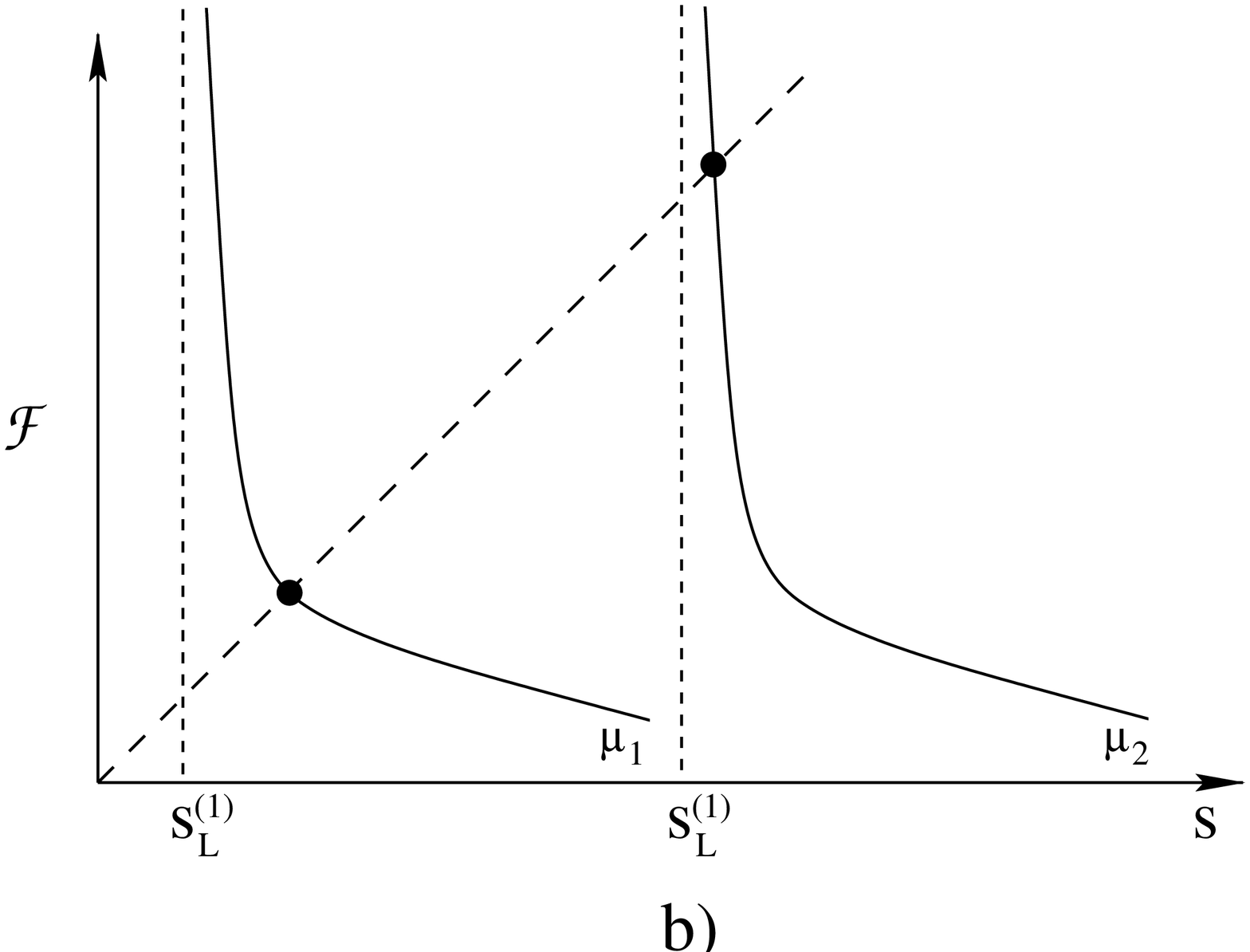}
}
\vskip-2mm\noindent{\footnotesize
Fig. 1.  Schematic view of singular points of the Isobaric Partition, Eq. (\ref{five}),
at $T < T_c$ (a) and $T > T_c$ (b).
Full lines show ${\cal F}(s,T,\mu)$ as a function of $s$ at fixed $T$ and $\mu$,
$\mu_1 < \mu_2 < \mu_3 < \mu_4$.
Dots and asterisks indicate the simple poles ($s_g$) and the singularity of
function ${\cal F}$ it-self ($s_l$).
At $\mu_3 = \mu^*(T)$ the two singular points coincide signaling a phase transition.
}
\label{fig:one}
\end{figure}


\newpage

\begin{figure}
\includegraphics[width=8.4cm]{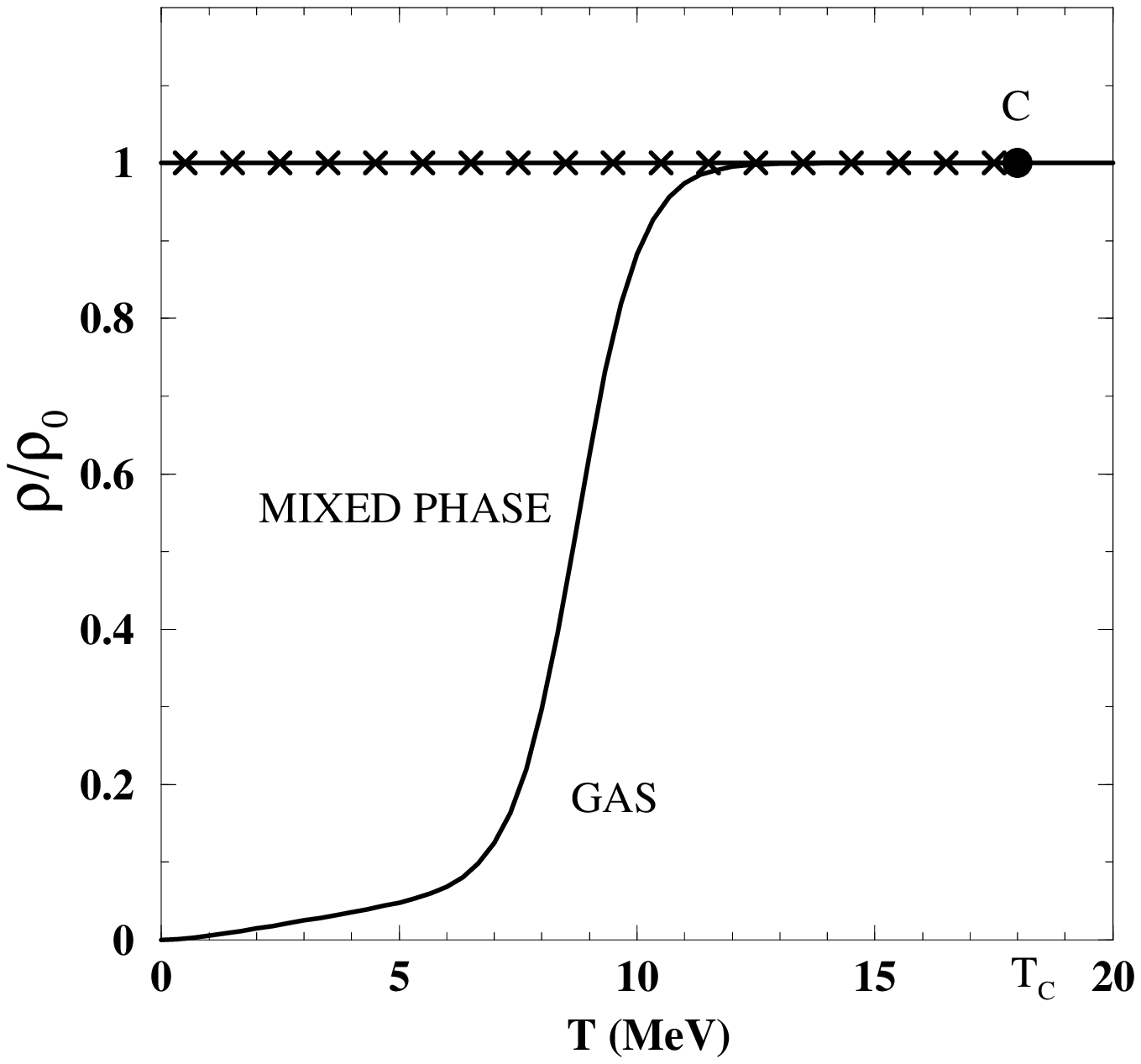}

\vspace*{-0.7cm}

\includegraphics[width=8.4cm]{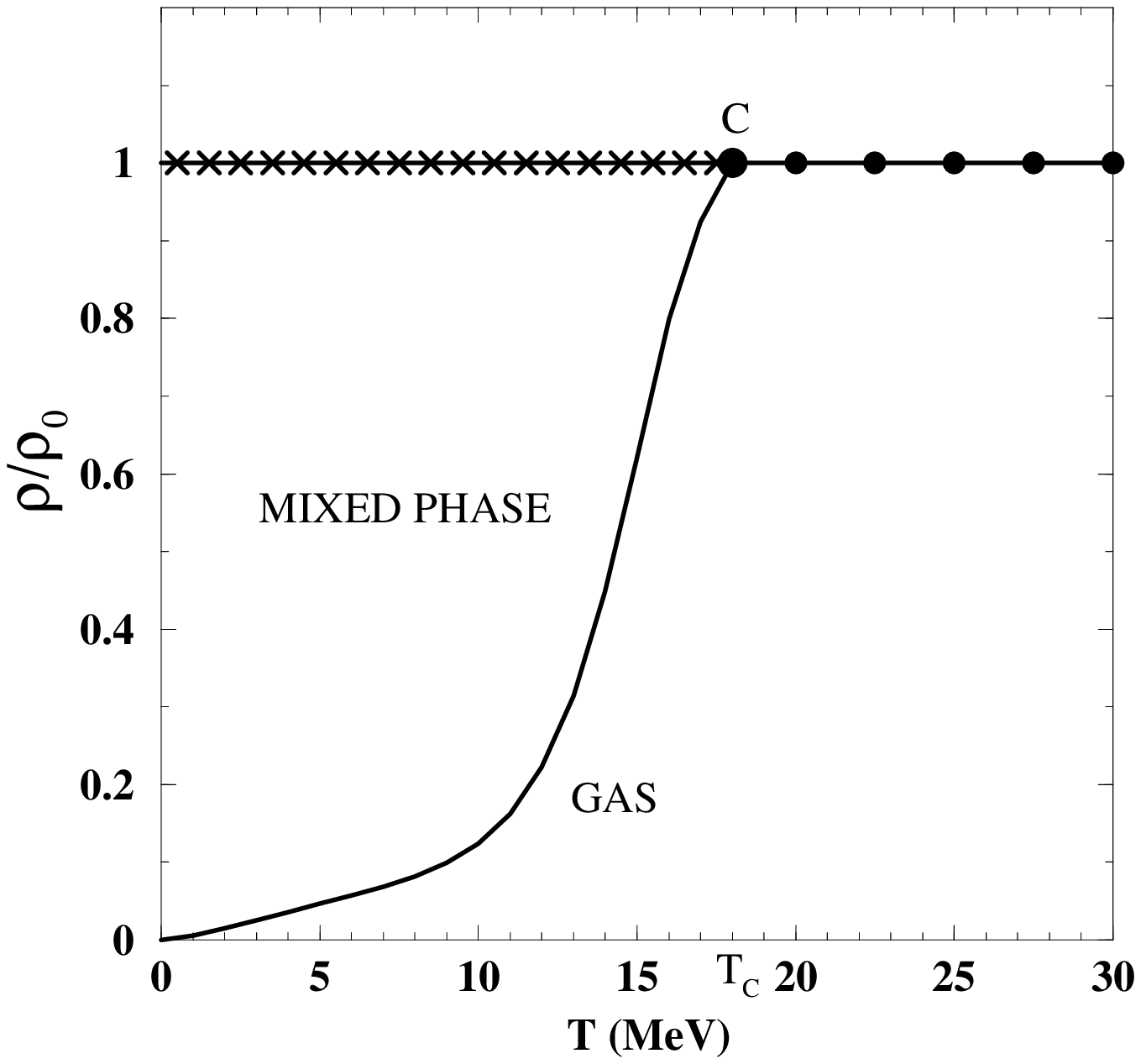}

\vspace*{-0.7cm}

\includegraphics[width=8.4cm]{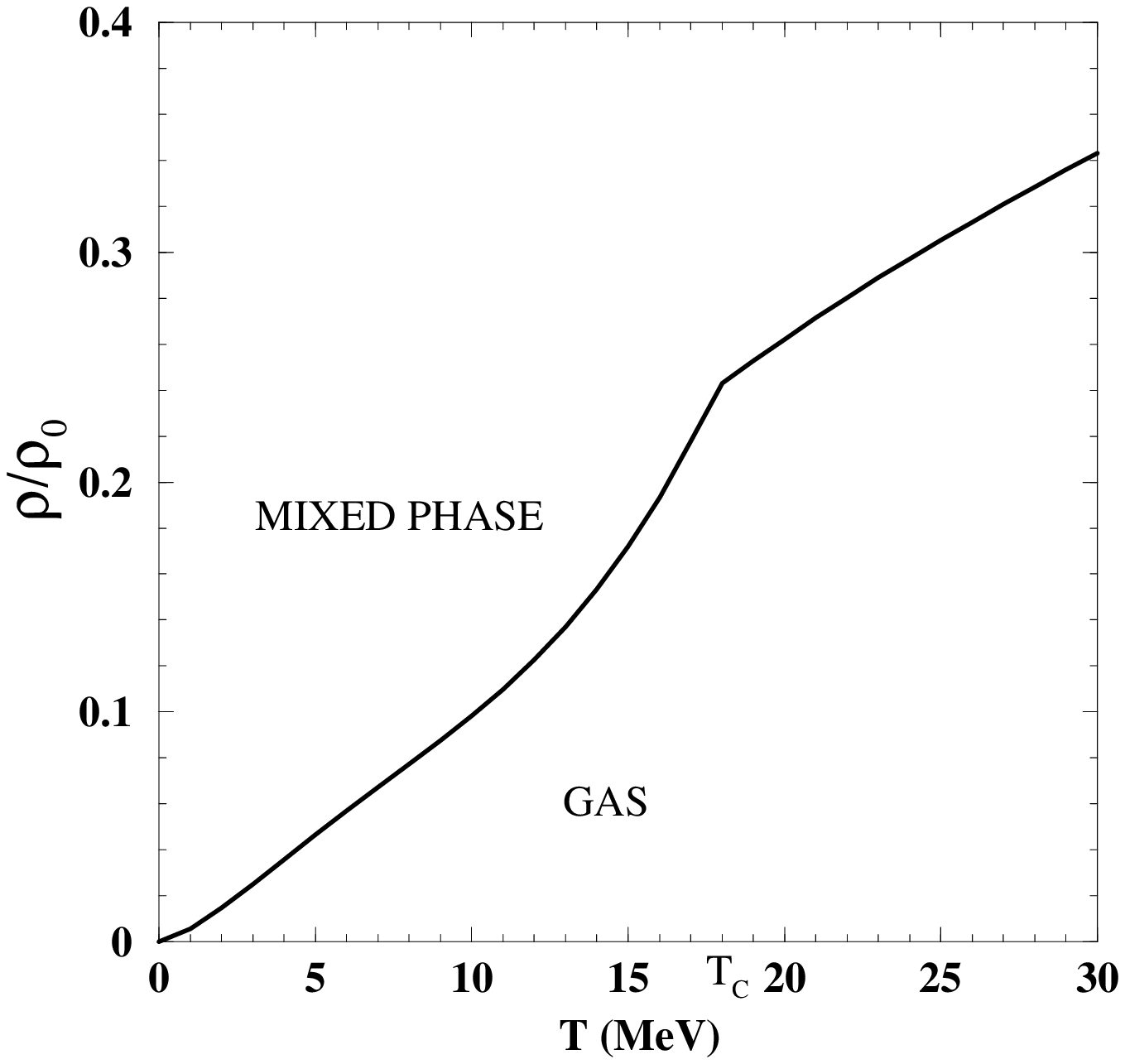}


\vskip-2mm\noindent{\footnotesize
Fig. 2. 
Phase diagram in  $T-\rho$ 
plane for $\tau = -1.5$ (upper panel),  $\tau =  1.1$ (middle panel) and  $\tau =  2.1$ (lower panel).
The mixed phase is represented by
the extended region.
Liquid phase (shown by crosses) exists at density $\rho = \rho_{\rm o}$.
Point $C$ in the upper panel  is the critical point, whereas in the middle panel it is the tricritical point.
For $\tau > 2$ (lower  panel) the PT exists for all temperatures $T \ge 0$.
}
\label{fig:two}
\end{figure}


\newpage

\clearpage

\begin{figure}
\includegraphics[width=8.4cm]{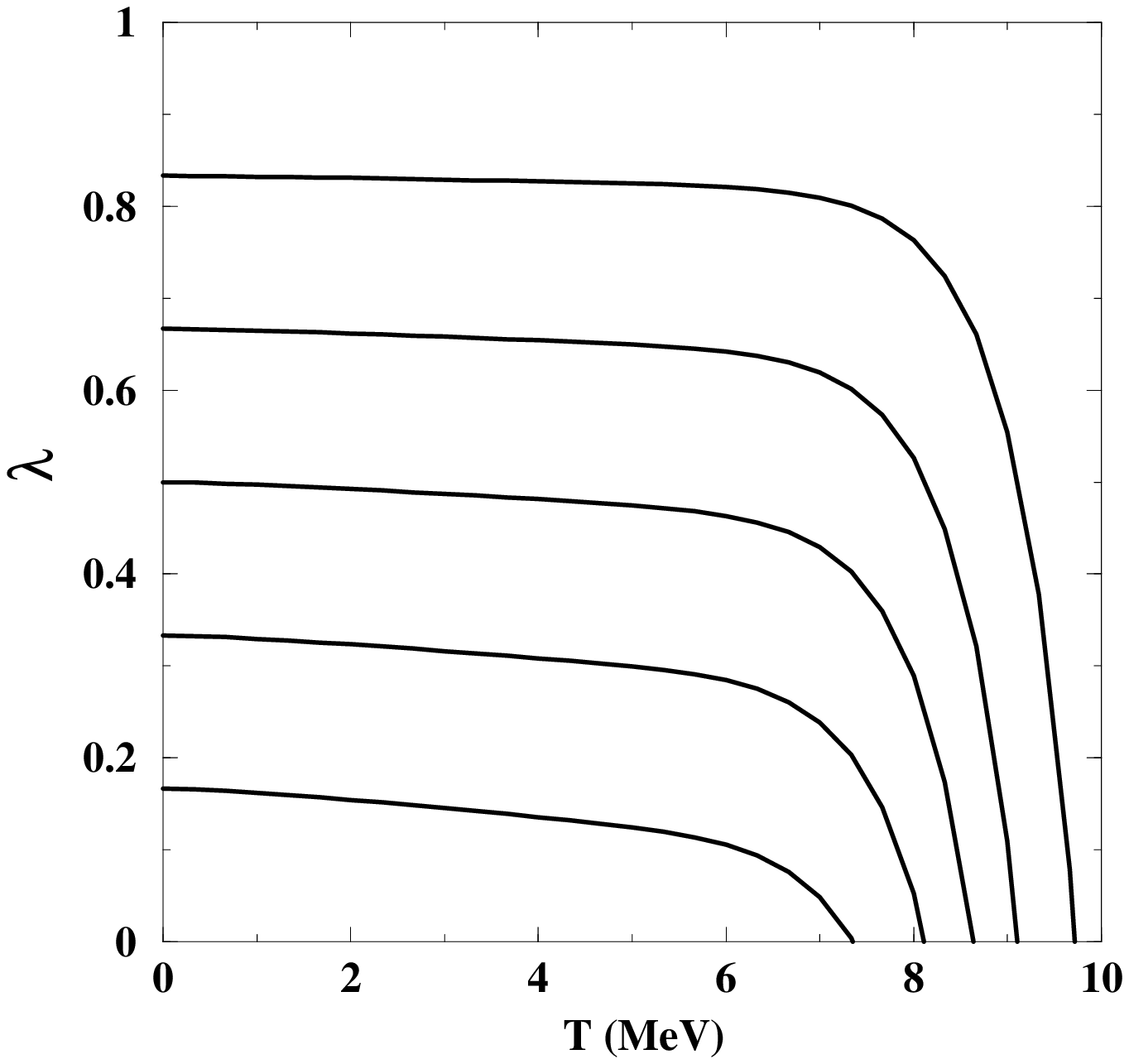}

\vspace*{-1.7cm}

\hspace*{-3.0cm}

\includegraphics[width=8.4cm]{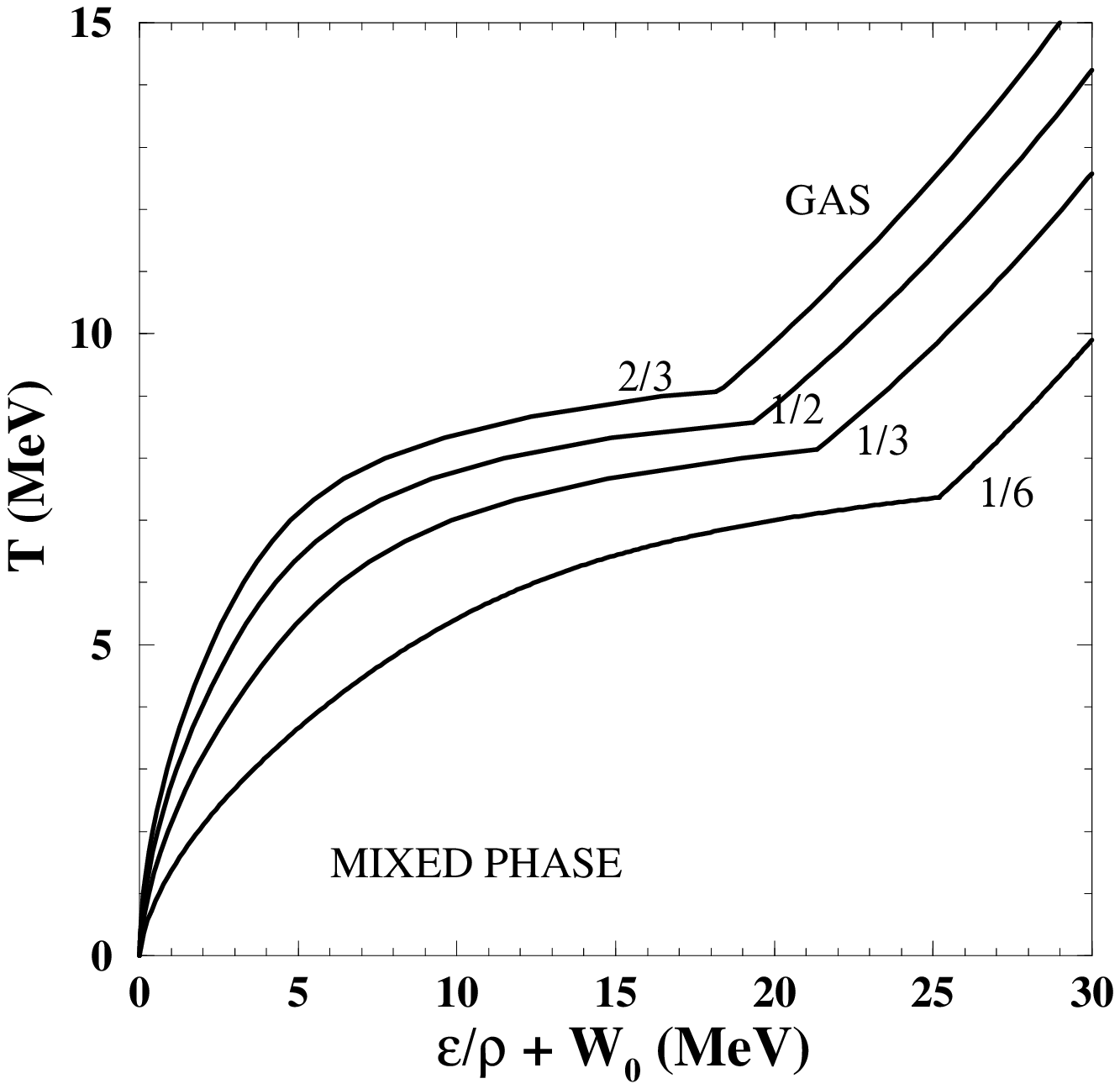}

\hspace*{-5.0cm}

\vspace*{-1.7cm}

\includegraphics[width=8.4cm]{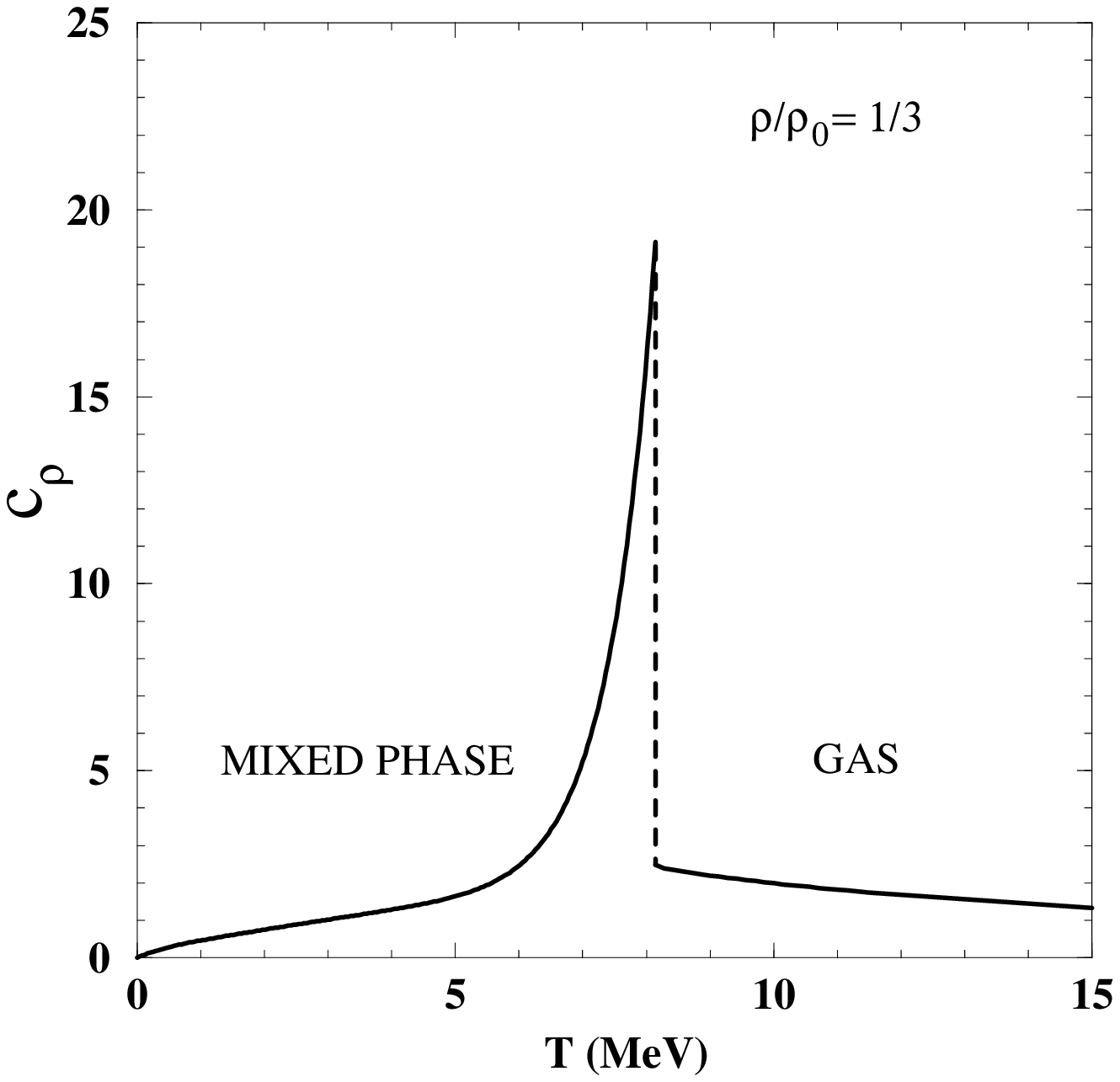}

\vspace*{-0.5cm}

\vskip-2mm\noindent{\footnotesize
Fig. 3. 
{\bf Upper panel:} Volume fraction $\lambda(T)$ of the liquid
inside the mixed phase is
shown as a function of temperature
for fixed nucleon densities ${\rho}/{\rho_{\rm o} } = 1/6, 1/3, 1/2, 2/3, 5/6$
(from bottom to top) and  $\tau = -1.5$.  \hfill \newline
{\bf Middle panel:} 
Temperature as a function of energy density per nucleon
(caloric curve)
is shown for fixed nucleon densities ${\rho}/{\rho_{\rm o} } = 1/6, 1/3,
1/2, 2/3$ and $\tau = -1.5$. {  Note that the shape  of the model caloric curves is very similar
to the experimental finding \cite{Natowitz:02}, although  our estimates for the excitation 
energy is somewhat larger due to oversimplified interaction. For  a quantitative comparison between 
the simplified SMM the full SMM interaction should be accounted for. 
}
 \newline
{\bf Lower panel:} 
Specific heat per nucleon as a function of temperature
at fixed nucleon density ${\rho}/{\rho_{\rm o} } = 1/3$. The dashed line
shows the finite discontinuity of $c_{\rho}(T)$
at the boundary of the mixed and gaseous phases for $\tau = -1.5$. }
\label{fig:three}
\end{figure}


\newpage

\begin{figure}[ht]

{

\includegraphics[width=7.6cm,height=6.0cm]{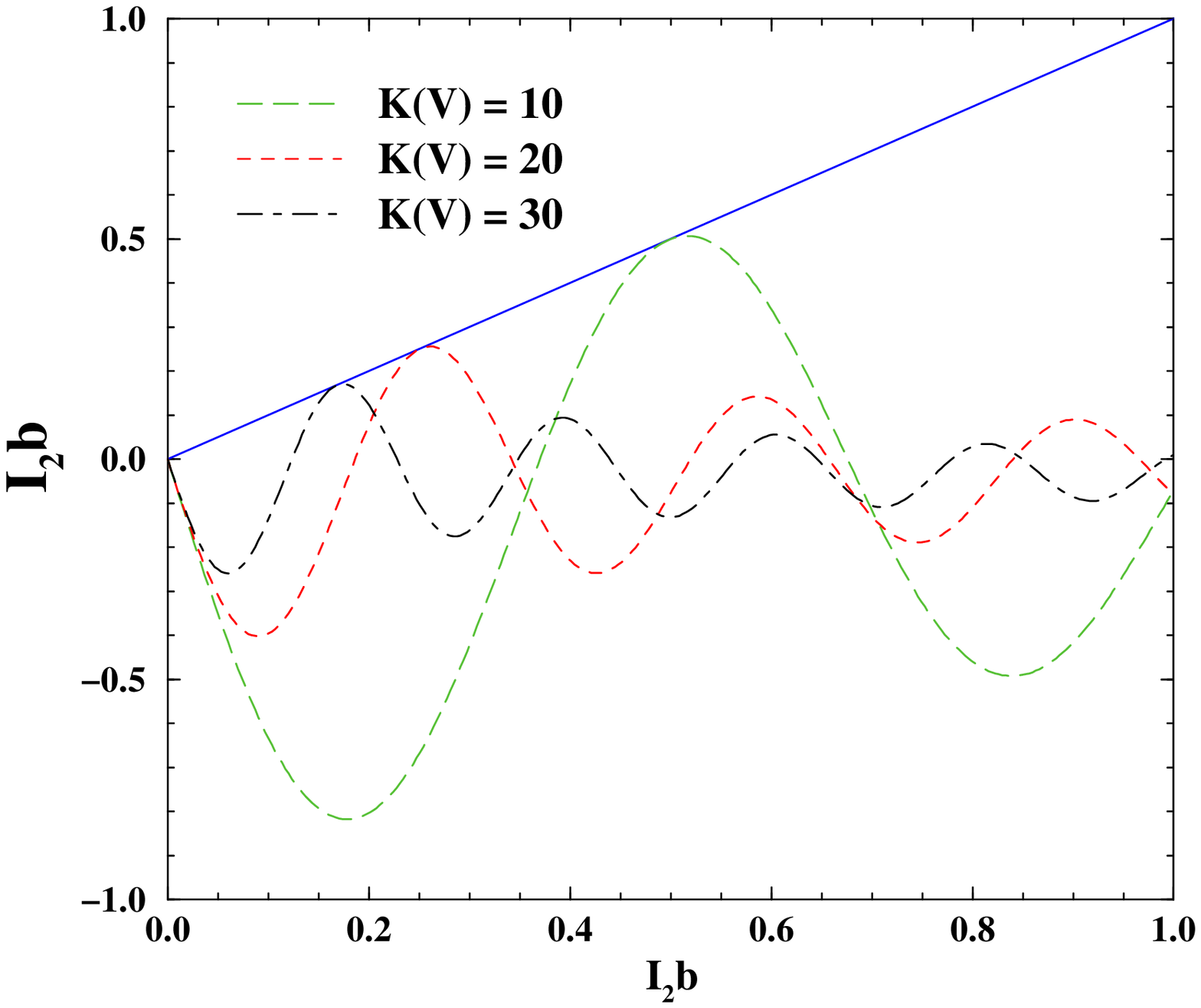}

\includegraphics[width=7.6cm,height=6.0cm]{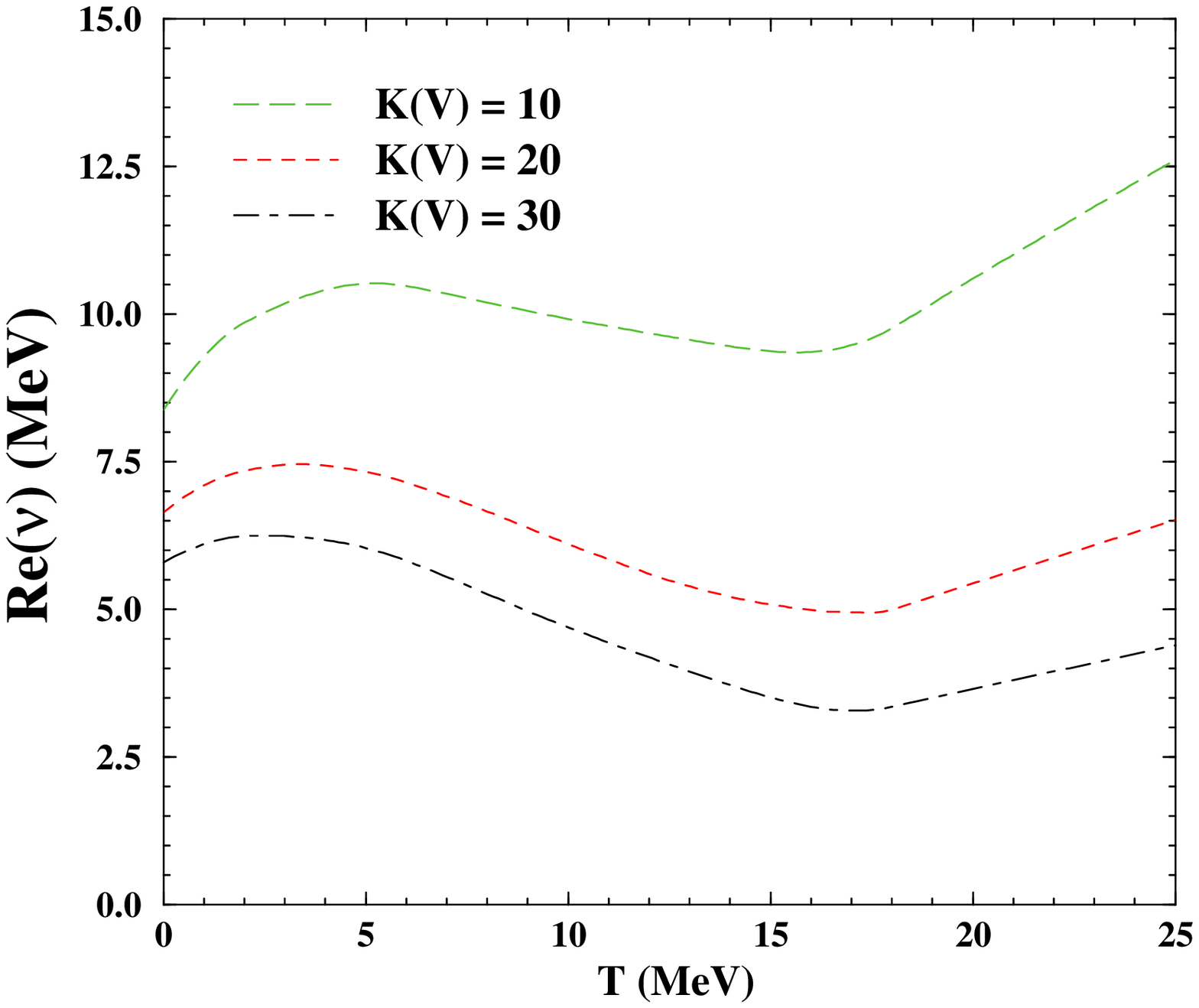} 
}

\vskip-2mm\noindent{\footnotesize
Fig. 4. 
{\bf Upper panel:} A graphical solution of Eq. (\ref{twelve}) for $T = 10$ MeV and $\tau = 1.825$.
The l.h.s. (straight line) and  r.h.s. of Eq. (\ref{twelve}) (all dashed curves) are shown
as the function of
dimensionless parameter $I_1\,b$ for the three values of the largest fragment size $K(V)$.
The intersection point at $(0;\,0)$ corresponds to a real root of Eq. (\ref{ten}).
Each tangent point with the straight line generates  two complex  roots of (\ref{ten}). \newline
{\bf Lower panel:}  Each curve separates  the $T-Re(\nu_n)$ region of one real root of Eq. (\ref{ten})
(below the curve), three complex roots (at the curve) and four and more roots (above the curve)
for three values of $K(V)$ and the same parameters as in the upper panel.
}
  \label{fig4}
\end{figure}


\begin{thebibliography}{99}


\bibitem{Bondorf:95}
J. P. Bondorf {\it et al.},
 Phys. Rep.  {\bf 257}, 131 (1995).

\bibitem{Gross:97}
D. H. E. Gross, Phys. Rep. {\bf 279}, 119 (1997).

\bibitem{Moretto:97}
L. G. Moretto {\it et al.},  Phys. Rep.  {\bf 287}, 249 (1997).

\bibitem{LeeYang}
%
C. N. Yang and T. D. Lee, Phys. Rev. {\bf 87}, 404 (1952).

\bibitem{Chomaz:03}
%
Ph. Chomaz and F. Gulminelli, Physica {\bf A 330}, 451 (2003).

\bibitem{Hill}
%
T. L. Hill, {\it Thermodynamics of Small Systems}, Dover Publications, N. Y., 1994.

\bibitem{Randrup:04}
%
P. Chomaz, M. Colonna and J. Randrup, Phys. Rep.  {\bf 389}, 263 (2004). 

\bibitem{Negheat:1}
%
F. Gulminelli  {\it et al.,}  Phys. Rev. Lett. {\bf 82}, 1402 (1999).


\bibitem{Negheat:2}
%
M. D`Agostino  {\it et  al.,}  Phys. Lett. {\bf B 473}, 219 (2000).


\bibitem{Bimodality:1} 
%
Ph. Chomaz, F. Gulminelli and V. Duflot, Phys. Rev.  {\bf E 64}, 046114 (2001).

\bibitem{Negheat:3}
%
L. G. Moretto {\it et al.},  Phys. Rev.  {\bf  C 66}, 041601(R)  (2002).

\newpage

\bibitem{Bimodality:2} 
\underline{L. G. Moretto,} J. B. Elliott and L. W. Phair,
{\bf  ``Mesoscopy and Thermodynamics''},  talk  given at the conference 
{\it ``World Consensus Initiative III''}, Texas A \& M University, College Station,
Texas, USA, February 11-17, 2005.
({see \bf http://cyclotron.tamu.edu/wci3/newer/chapVI\_4.pdf})



\bibitem{QM:04}
%
M. Gyulassy, 
Lect. Notes Phys.  {\bf 583}, 37 (2002).



\bibitem{Gupta:98}
S. Das Gupta and A.Z. Mekjian,   Phys. Rev.  {\bf C 57}, 1361 (1998).

\bibitem{Gupta:99}
S. Das Gupta, A. Majumder, S. Pratt and A. Mekjian,
arXiv:nucl-th/9903007.


\bibitem{Bugaev:00}
K. A. Bugaev, M. I. Gorenstein, I. N. Mishustin and W. Greiner,
Phys. Rev. {\bf C 62},  044320 (2000);
arXiv:nucl-th/0007062.

\bibitem{Bugaev:01}
K. A. Bugaev,  M. I. Gorenstein, I. N. Mishustin and W. Greiner,
Phys. Lett. {\bf B 498}, 144  (2001);
arXiv:nucl-th/0103075.

\bibitem{Reuter:01}
P. T. Reuter and K. A. Bugaev,
Phys.\ Lett. {\bf B 517}, 233  (2001).

\bibitem{Fisher:67}
%
M. E. Fisher, Physics {\bf 3}, 255  (1967).


\bibitem{ISIS}
%
L. Beaulieu  {\it et al.,}  Phys. Lett. {\bf B 463}, 159 (1999).


\bibitem{EOS:00}
J. B. Elliott {\it et al.,} (The EOS Collaboration),
Phys. Rev. {\bf C 62},  064603 (2000).

\bibitem{Karnaukhov:tau}
%
V. A.  Karnaukhov {\it et al.,} Phys. Rev. {\bf C 67}, 011601R (2003);
N. Buyukcizmeci, R. Ogul and A. S. Botvina,  arXiv:nucl-th/0506017 and references therein. 


\bibitem{Elliott:05wci}
%
the most recent results can be found in  the review 
J. B. Elliott, K. A. Bugaev,   L. G. Moretto and  L. Phair,
 arXiv:nucl-ex/0608022 (2006) 36 p.




\bibitem{Elliott:02}
J. B. Elliott {\it et al.,} Phys. Rev. Lett. {\bf 88}, 042701 (2002).

\bibitem{Elliott:03}
J. B. Elliott {\it et al.,} Phys. Rev.  {\bf C 67}, 024609 (2003).


\bibitem{Precomplement} 
%
J. B. Elliott, L. G. Moretto and L. Phair,  Phys. Rev. {\bf C 71}, 024607 (2005). 

\bibitem{Complement} 
%
L. G. Moretto {\it et al.},  Phys. Rev. Lett. {\bf 94},  202701 (2005).


\bibitem{Bugaev:04a}
K. A. Bugaev,
Acta. Phys. Pol.  {\bf B 36},  3083 (2005). 
%



\bibitem{Bugaev:04b}
K. A. Bugaev, L. Phair and J. B. Elliott,
Phys. Rev.  {\bf E 72},  047106 (2005);
K. A. Bugaev  and J. B. Elliott,
arXiv:nucl-th/0501080.

\bibitem{Bugaev:05}
%
K. A. Bugaev ,  in preparation. 


\bibitem{Goren:81}
M. I. Gorenstein, V. K. Petrov and G. M. Zinovjev, Phys. Lett.
{\bf B 106},  327 (1981).

\bibitem{Natowitz:02}
%
J. Natowitz   {\it et al.},   Phys. Rev.  {\bf C 65}, 034618  (2002). 

\bibitem{Elliott:01}
%
J. B. Elliott {\it et al.}  [ISiS Collaboration],
Phys.\ Rev.\ Lett.\  {\bf 88},  042701 (2002).



\bibitem{Ya:64}
C. N. Yang and C. P. Yang, Phys. Rev. Lett. {\bf 13},   303 (1964).

\bibitem{Fi:70}
M. E. Fisher and B. U. Felderhof, Ann. of Phys. {\bf 58},  217 (1970).

\bibitem{Fi:64}
M. E. Fisher, J. Math. Phys. {\bf 5},  944 (1964).

\bibitem{Gri:65}
R. B. Griffiths, J. Chem. Phys. {\bf 43},   1958 (1965).

\bibitem{Li:66}
D. A. Liberman, J. Chem. Phys. {\bf 44},  419 (1966).

\bibitem{Eos:94}
M. L. Gilkes {\it et al.},
Phys. Rev. Lett. {\bf 73},  1590  (1994).

\bibitem{Bau:94}
%
W. Bauer and W. A. Friedman,
arXiv:nucl-th/9411012.

\bibitem{Hu}
K. Huang, {\it Statistical Mechanics}, J. Wiley, New York, 1987.

\bibitem{Zi:98}
R. Guida and J. Zinn-Justin, J. Phys. Math. Gen. {\bf 31}, 8103 (1998). 


\bibitem{Pan:83}
A. D. Panagiotou {\it et al.}, Phys. Rev. Lett. {\bf 52},  496 (1983).


\bibitem{El:00}
J. B. Elliott and A. S. Hirsch,
Phys. Rev. {\bf C 61}, 054605 (2000).


\bibitem{CSMM}
C.~B.~Das, S.~Das Gupta and A.~Z.~Mekjian,
Phys.\ Rev. {\bf C 68},   031601 (2003).

\bibitem{Bugaev:05csmm}
%
K. A. Bugaev, arXiv:nucl-th/0507028.

\bibitem{Feynmann}
%
R. P. Feynman, {\it ``Statistical Mechanics'',} Westview Press, Oxford, 1998.


\bibitem{Goren:05}
%
M. I. Gorenstein, M. Gazdzicki and W. Greiner,
%
{\sc Check Ref!!!} {arXiv: nucl-th/0505050} and references therein.


\bibitem{Hgas}
%
G. D. Yen and M. I. Gorenstein, Phys. Rev.  {\bf C  59},  2788 (1999);
P. Braun-Munzinger, I. Heppe and J. Stachel, Phys. Lett. B {\bf
465}, 15 (1999).

\bibitem{BagModel}
%
A. Chodos {\it et al.,}  Phys. Rev.  {\bf D 9},  3471 (1974). 

\bibitem{Goren:82}
%
M. I. Gorenstein, G. M. Zinovjev, V. K. Petrov, and V. P. Shelest, 
Teor. Mat. Fiz. (Russ) {\bf 52}), 346  (1982).

\bibitem{HThermostat:1}
%
L. G. Moretto,  K. A. Bugaev, J. B. Elliott and L. Phair,
LBNL preprint {\bf  56898;}
arXiv: nucl-th/0504010  (to appear in  Euro. Phys. Lett.) .


\bibitem{HThermostat:2}
%
K. A. Bugaev,  J. B. Elliott, L. G. Moretto and  L. Phair,
arXiv: hep-ph/0504011.


\bibitem{Ising:clust}
%
C. M. Mader {\it et al.},
Phys. Rev.  {\bf C 68},,  064601 (2003).


\bibitem{Dillmann}
%
A. Dillmann and G. E. A. Meier, J. Chem. Phys. {\bf 94},  3872 (1991).

\bibitem{Kiang}
%
C. S. Kiang, Phys. Rev. Lett. {\bf 24},  47 (1970).


\bibitem{Percolation}
%
D. Stauffer and A. Aharony, ``Introduction to Percolation", Taylor and Francis, Philadelphia, 2001. 

\bibitem{Heinz:05}
%
U. Heinz, arXiv:nucl-th/0504011. 

\bibitem{Bugaev:96} 
%
K. A. Bugaev,  
Nucl. Phys. {\bf A 606}, 559 (1996);
%
K. A. Bugaev and  M. I. Gorenstein, 
arXiv:nucl-th/9903072. 

\bibitem{Bugaev:99} 
K. A. Bugaev, M. I. Gorenstein and W. Greiner, 
J. Phys. {\bf G 25}, 2147 (1999);  Heavy Ion Phys. {\bf 10}, 333 (1999). 

\bibitem{Rischke1}
D. H. Rischke,  ``Fluid Dynamics for Reativistic Nuclear Collisions'', 
the proceedings of 11th Chris Engelbrecht Summer School in Theoretical 
Physics: Hadrons in Dense Matter and Hadrosynthesis, 
Cape Town, South Africa, 4-13 Feb 1998.
In *Cape Town 1998, Hadrons in dense matter and hadrosynthesis* 21-70;
arXiv: nucl-th/9809044. 


\bibitem{BD:00}
S. A. Bass and A. Dumitru, Phys. Rev. {\bf C 61},  064909 (2000).

\bibitem{TLS:01}
D. Teaney, J. Lauret and E. V. Shuryak,
Phys. Rev. Lett. {\bf 86},  4783 (2001);
%
arXiv:nucl--th/0110037.

\bibitem{Bugaev:02HC}
%
K. A. Bugaev, Phys. Rev. Lett. {\bf 90}, 252301 (2003).

\bibitem{Bugaev:04HC}
%
K.~A.~Bugaev,
Phys. Rev.  {\bf C 70}, 034903 (2004).



\end{thebibliography}
\end{document}